\documentclass[10pt,twocolumn,letterpaper]{article}

\usepackage{cvpr}              

\usepackage{microtype}

\renewcommand{\paragraph}[1]{\vspace{.5em}\noindent\textbf{#1.}}

\usepackage{soul}
\setuldepth{foobar}
\usepackage{algorithm}
\usepackage{algorithmic}
\usepackage{multirow}
\usepackage{subcaption}
\usepackage{booktabs}
\usepackage{amsmath} 
\usepackage{amssymb}
\usepackage{makecell}

\definecolor{cvprblue}{rgb}{0.21,0.49,0.74}
\usepackage[pagebackref,breaklinks,colorlinks,allcolors=cvprblue]{hyperref}

\def\paperID{35713} 
\def\confName{CVPR}
\def\confYear{2026}

\title{CLIP-Inspector: Model-Level Backdoor Detection for Prompt-Tuned CLIP via OOD Trigger Inversion}

\author{Akshit Jindal$^1$\thanks{Corresponding author: akshitj@iiitd.ac.in} \quad Saket Anand$^1$ \quad Chetan Arora$^2$ \quad Vikram Goyal$^1$\\[0.3em]
$^1$IIIT Delhi $^2$IIT Delhi\\
}

\begin{document}
\maketitle
\begin{abstract}
Organisations with limited data and computational resources increasingly outsource model training to Machine Learning as a Service (MLaaS) providers, who adapt vision-language models (VLMs) such as CLIP to downstream tasks via prompt tuning rather than training from scratch. This semi-honest setting creates a security risk where a malicious provider can follow the prompt-tuning protocol yet implant a backdoor, forcing triggered inputs to be classified into an attacker-chosen class, even for out-of-distribution (OOD) data. Such backdoors leave encoders untouched, making them undetectable to existing methods that focus on encoder corruption. Other data-level methods that sanitize data before training or during inference, also fail to answer the critical question, \textbf{``Is the delivered model backdoored or not?"}
To address this model-level verification problem, we introduce CLIP-Inspector (CI), a backdoor detection method designed for prompt-tuned CLIP models. Assuming white-box access to the delivered model and a pool of unlabeled OOD images, CI reconstructs possible triggers for each class to determine if the model exhibits backdoor behaviour or not. Additionally, we demonstrate that using CI’s reconstructed trigger for fine-tuning on correctly labeled triggered inputs enables us to re-align the model and reduce backdoor effectiveness.
Through extensive experiments across ten datasets and four backdoor attacks, we demonstrate that CI can reconstruct effective triggers in a single epoch using only 1,000 OOD images, achieving a 94\% detection accuracy (47/50 models). Compared to adapted trigger-inversion baselines, CI yields a markedly higher AUROC score (0.973 vs 0.495/0.687), thus enabling the vetting and post-hoc repair of prompt-tuned CLIP models to ensure safe deployment.
\end{abstract}    
\section{Introduction}

Vision-language foundation models (VLMs), particularly Contrastive Language-Image Pre-training (CLIP)~\cite{radford2021learning}, have reshaped modern computer vision by enabling strong zero-shot and few-shot performance across diverse tasks. In practice, however, many organisations lack the data, compute, or expertise to adapt these models for their tasks. Instead, they outsource training to Machine Learning as a Service (MLaaS) providers, supplying them with proprietary data and receiving a fine-tuned CLIP model optimized for their specific needs. The prompt-tuning mechanism, Conditional Context Optimization (CoCoOp)~\cite{zhou2022conditional}, is particularly effective in this setting, as a small image-conditioned meta-network can be efficiently trained to generate instance-specific context tokens, yielding highly accurate classifiers.

\begin{figure}
  \centering
  \includegraphics[height=0.62\linewidth,width=\linewidth]{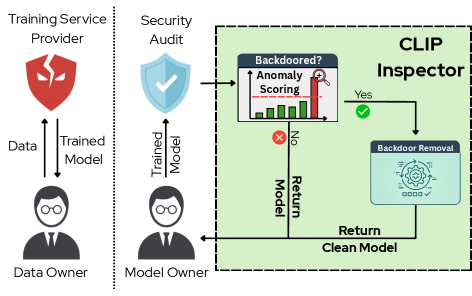}
  \caption{Threat model and audit workflow. A semi-honest provider can implant a backdoor into the CLIP model while following the specified prompt-tuning protocol. A security auditor uses our method to determine whether the model is backdoored and, if necessary, performs a light repair using the reconstructed trigger.}
  \label{fig:teaser_fig}
\end{figure}

This outsourced, semi-honest pipeline introduces a concrete security threat. A malicious provider can follow the prescribed prompt-tuning protocol and deliver a high-performance model while also secretly implanting backdoors. Any input stamped with an imperceptible trigger would be classified into an attacker-chosen class, even when it is far out of distribution (OOD). Once the model is deployed in a safety- or security-critical system (e.g., wildlife monitoring or autonomous access control), the backdoor can be exploited to bypass detection or redirect decisions. This necessitates model security verification prior to deployment. Typically, the model owner sees only the delivered model weights and has no way of identifying malicious behaviour. As illustrated in Figure~\ref{fig:teaser_fig}, a realistic defence workflow is to route the delivered model through an independent security audit service to answer the critical question: \emph{is this model backdoored or not?}

\begin{figure*}[t]
    \centering
    \begin{subfigure}[b]{0.59\textwidth}
        \includegraphics[height=4.3cm,width=10cm]{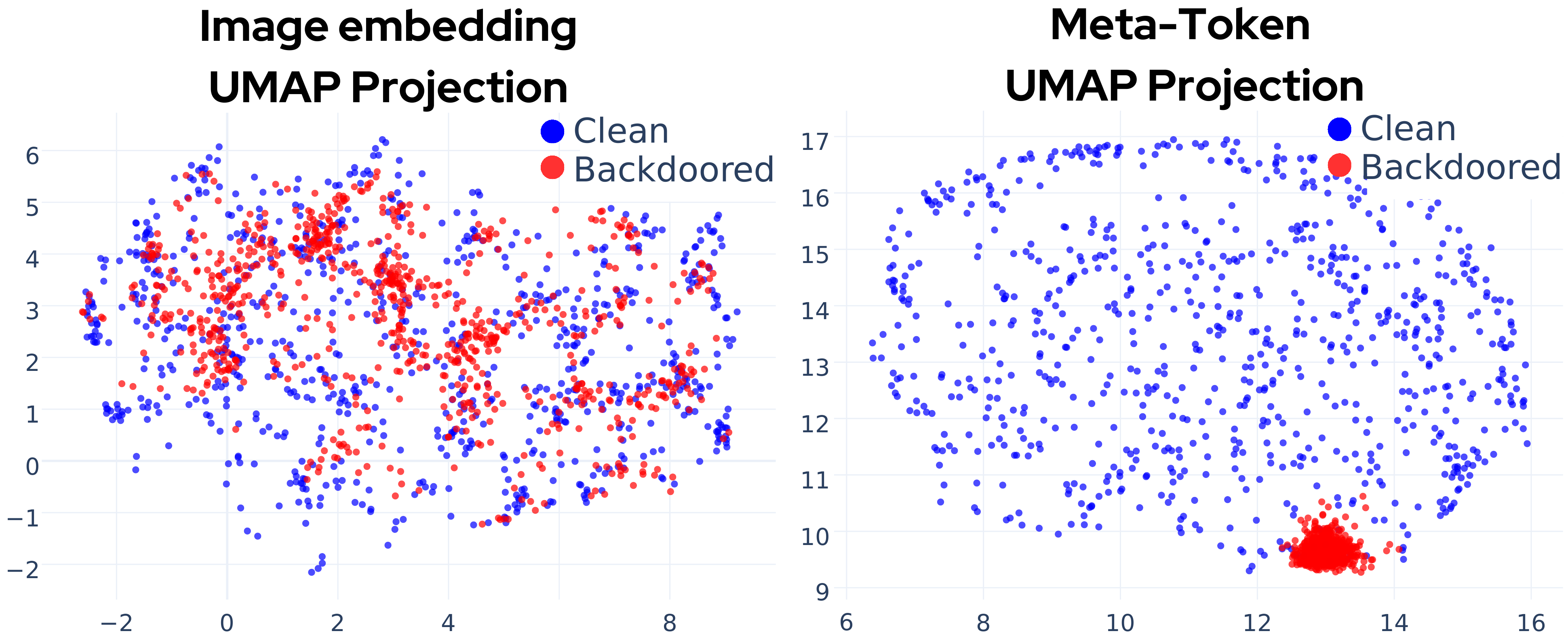}
        \caption{}
        \label{fig:logit_dist_intro}
    \end{subfigure}
    \hspace{0.03cm}
    \begin{subfigure}[b]{0.39\textwidth}
        \includegraphics[height=4.3cm,width=6.5cm]{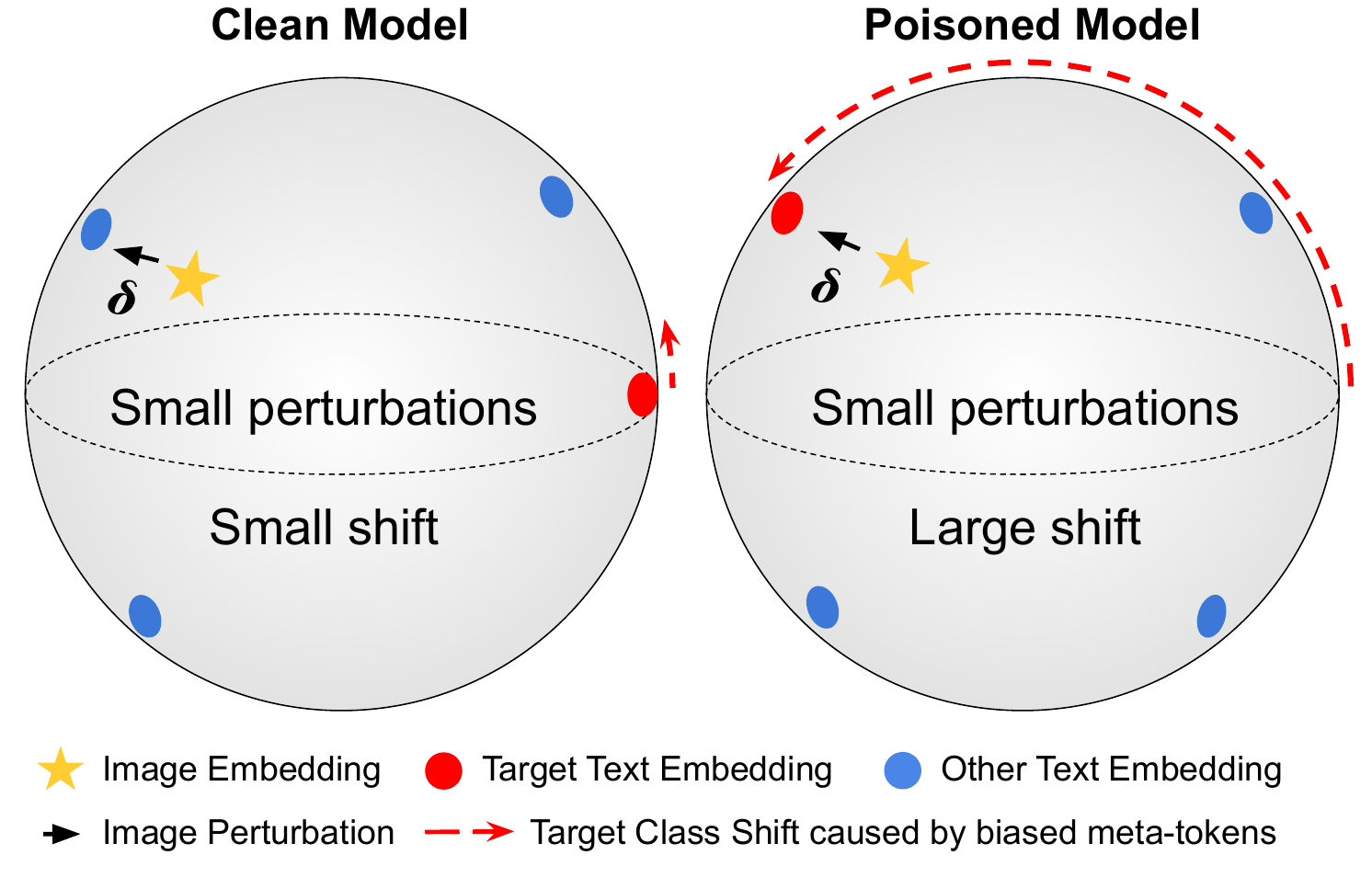}
        \caption{}
        \label{fig:image_embed_intro}
    \end{subfigure}
    \caption{(a) Image-space embeddings for clean (blue) and poisoned (red) OOD inputs substantially overlap, while meta-tokens for poisoned inputs form tight clusters. (b) For a poisoned model, even a slight image perturbation can cause substantial changes in the targeted text embedding, resulting in the backdoor effect. In contrast, for a clean model, the shift is negligible.}
    \label{fig:intro_fig}
\end{figure*}

In the CoCoOp prompt-tuning mechanism, training happens without explicit textual prompts. Recently, BadCLIP ~\cite{bai2024badclip} demonstrated that a backdoor can be implanted by jointly optimizing an imperceptible image-wide perturbation along with the model. The perturbation only slightly affects the image embedding but leads to significant bias in meta-net outputs. Figure~\ref{fig:intro_fig} shows that while poisoned image embeddings largely overlap with clean ones, the meta-tokens for poisoned inputs collapse to a single cluster. Subsequently, the biased meta-tokens significantly reorient the text embeddings to achieve the desired backdoor behaviour. In addition to BadCLIP, we show that existing image-space backdoor attacks \cite{nguyen2021wanet,blended2017,siba2024backdoor} can also be adapted to the prompt-tuning setting, making prompt-tuning especially vulnerable to backdoor attacks.

As CLIP's encoders remain frozen during prompt tuning, the backdoor functions entirely through the meta-net. This mechanism breaks the assumptions underlying many model-level detection methods. Existing methods either assume that (a) the trigger is sparse \cite{wang2019neural, guo2019taborhighlyaccurateapproach, pixelbackdoor2022} or (b) the encoders are poisoned \cite{wang2023unicornunifiedbackdoortrigger, wang2022featuREdetect, feng2023detecting, sur2023tijo, exray}. Methods that assume sparsity and use $L_1$ norms inflate a few pixels to extreme values and fail to reconstruct imperceptible triggers. Methods that analyze encoder behaviour \cite{ulpattern, practicaltrojdetecteccv, seerdetection} cannot differentiate between a clean and a backdoored prompt-tuned model. Moreover, as our goal is pre-deployment screening, test-time defenses \cite{gao2019strip,liu2023detectingteco,guo2023scaleup,sun2023mask,niu2024bdetclip} are directly inapplicable. This necessitates the need for a specialized detection method for the prompt-tuning scenario. 

In this paper, we present \textbf{CLIP-Inspector (CI)}, a model-level backdoor detector designed for prompt-tuned CLIP models. CI probes a delivered model using only unlabeled out-of-distribution images, leveraging the strong OOD generalization capabilities of prompt-tuned backdoors\cite{bai2024badclip}. For each candidate class, it reconstructs an imperceptible image-wide perturbation with a margin-based objective on the model’s logits, then evaluates both the attack success rate and the optimization loss on held-out OOD data. Aggregating these signals yields a simple behaviour-driven anomaly score that allows us to simultaneously assess whether the model is backdoored and determine the target class. Although our method is designed for prompt-tuned CLIP, we also evaluate encoder-level backdoors (no meta-net) and observe that CI is able to detect them without any changes. 

Beyond detection, we demonstrate that CI’s reconstructed perturbations are not merely diagnostic artifacts and can be utilized for model purification. Fine-tuning the prompt-tuned model on triggered inputs with their ground-truth labels significantly reduces the backdoor ASR while preserving clean accuracy. This matches a realistic audit workflow where an organisation receives weights from an MLaaS provider, submits them to an independent security service with white-box access for vetting, and, if needed, requests a quick post-hoc repair before deployment (see Figure \ref{fig:teaser_fig}). 

Our main contributions are:
\begin{itemize}
    \item We formulate and study \emph{model-level} backdoor detection for prompt-tuned CLIP classifiers in an outsourced, semi-honest setting, where the attacker can manipulate model behaviour while keeping the encoders intact.
    \item We introduce \textbf{CLIP-Inspector}, an OOD trigger-inversion-based detector that reconstructs class-wise perturbations to yield a model-level clean-vs-backdoored decision and identify the target class without any in-distribution data.
    \item We conduct a comprehensive evaluation across ten datasets and four attacks (BadCLIP and three adapted attacks), significantly outperforming adapted trigger-inversion baselines (NC \cite{wang2019neural} and PixB \cite{pixelbackdoor2022}).
    \item We demonstrate that CI’s reconstructed perturbations are functionally similar to the attacker’s trigger and are capable of reducing ASR to $<10\%$ in a couple of fine-tuning steps while preserving ACC ($\pm1\%)$
\end{itemize}

\section{Background and Related Work}
CLIP or Contrastive Language Image Pretraining (\cite{radford2021learning}) is a learning paradigm that uses separate encoders to project image and text inputs into a shared embedding space such that semantically similar concepts are positioned together. Given an input image $x$, its feature representation is denoted as $f_I(x) \in \mathbb{R}^{512}$. The text encoder ($f_T$) takes as input a combination of textual context and class names, such as ``a photo of a [CLS],'' where [CLS] denotes the class name. Formally, given word embedding vectors for textual context tokens $V = [v_1, v_2, ..., v_N]^{\top} \in \mathbb{R}^{N \times e}$ and a class name embedding $c_i \in \mathbb{R}^e$ for class $i$, the probability of assigning $x$ to class $i$ is computed as:
\begin{equation}
  p(y = i|x) = \frac{\exp(\text{sim}(f_I(x), f_T(\{V, c_i\}))/\tau)}{\sum_{j=1}^{K} \exp(\text{sim}(f_I(x), f_T(\{V, c_j\}))/\tau)},
\end{equation}
where $\text{sim}(\cdot, \cdot)$ denotes cosine similarity, and $\tau$ is the learned temperature coefficient of CLIP. The CLIP model's zero-shot classification capabilities have made it a cornerstone for various vision-language tasks.

To adapt CLIP to specific downstream tasks with limited data, practitioners rely on efficient fine-tuning techniques, such as prompt-tuning or CoCoOp. CoCoOp employs a meta-net $h_\theta(\cdot)$ that generates image-specific tokens, which are then combined with learnable context vectors and class name embedding to produce prompts $\{h_{\theta}(x), c_i\}$, where $h_{\theta}(x) \in \mathbb{R}^{N \times e}$, $i = 1, 2, ..., K$, and K is the number of output classes. The corresponding prediction probability is thus computed as:
\begin{equation}
  \tilde{p}(y = i | x) = \frac{\exp(\text{sim}(f_I(x), f_T(\{h_{\theta}(x), c_i\}))/\tau)}{\sum_{j=1}^{K} \exp(\text{sim}(f_I(x), f_T(\{h_{\theta}(x), c_j\}))/\tau)}.
\end{equation}
For efficiency, $h(\cdot)$ is typically implemented as a two-layer fully connected network, and is the only component trained \textit{while the image and text encoders are kept frozen}.

Backdoor attacks seek to manipulate model behaviour via specialized perturbations of the input data. These perturbations can be either\textit{ patch-like} \cite{gu2019badnets, liu2018trojaning, nguyen2020inputaware, saha2019hidden, liang2024badclip} or\textit{ pervasive} \cite{li2021invisible, siba2024backdoor, blended2017, nguyen2021wanet}. Recently, BadCLIP \cite{bai2024badclip} demonstrated that backdoors can be injected during the CoCoOp process. They show that the meta-net can be trained to output biased tokens in the presence of an imperceptible input trigger ($\delta$). The model is trained to maximize $\tilde{p}(y = t | x + \delta) \,\forall x$, where t is the backdoor target. Their method achieves an ASR of greater than 99\% while maintaining clean accuracy and can even be used to adapt existing image-space backdoors to the prompt-tuning scenario, thus necessitating the need for a specialized detection method.

Backdoor detection is typically performed at the (i) input-level, (ii) dataset-level, or (iii) model-level. Input-level detection is aimed at identifying whether an incoming input image sample is poisoned or not at the inference stage \cite{gao2019strip, liu2023detectingteco, guo2023scaleup, niu2024bdetclip}, whereas dataset-level detection, also known as dataset purification, is conducted before model training to find triggers in the training data \cite{huangDetectingBackdoorSamples2025a, hou2024dede}. On the other hand, model-level detection \cite{wang2019neural,guo2019taborhighlyaccurateapproach, wang2023unicornunifiedbackdoortrigger,wang2022featuREdetect,pixelbackdoor2022, sur2023tijo} aims to determine whether a given model is backdoored or not. It is usually conducted via trigger inversion or encoder output analysis. We limit our discussion to model-level detection methods in this paper.

Existing detection methods make certain assumptions that do not hold for prompt-tuned backdoors.
\textbf{(I)} inversion methods that assume the trigger is sparse (small localized patch or limited to a few pixels) \cite{wang2019neural, guo2019taborhighlyaccurateapproach, tejankar_defending_2023, shen2021backdoorscanningkarm, pixelbackdoor2022} are likely to inflate a few pixels to extreme values and are thus incapable of generating imperceptible triggers. These methods are also known to converge to universal adversarial noise instead of intentionally embedded backdoors, leading to false positives \cite{grosse2022backdoor,le2024doubleeuspico}. 
\textbf{(II)} Feature-space detection methods \cite{wang2022featuREdetect, wang2023unicornunifiedbackdoortrigger} designed for unimodal CNNs assume that the image embedding fully governs classification, as they seek feature-space activations specific to backdoor behaviour. In a poisoned prompt-tuned CLIP model, the backdoor effect is achieved by realigning text embeddings, while the image embedding remains largely unaffected. 
\textbf{(III)} encoder-based methods assume (i) the pre-training data has been poisoned, and (ii) the encoder has been trained to produce poisoned embeddings. \cite{feng2023detecting, sur2023tijo,seerdetection}. 
Such methods are also ineffective because prompt-tuning does not modify the underlying encoders, and the training data is poisoned during prompt-tuning, not beforehand.
\textit{Overall, we argue that none of the existing methods would be able to detect prompt-tuned backdoor attacks on CLIP models.}

\section{Methodology}
\begin{figure*}[t]
  \centering
  \includegraphics[height=0.33\linewidth,width=\linewidth]{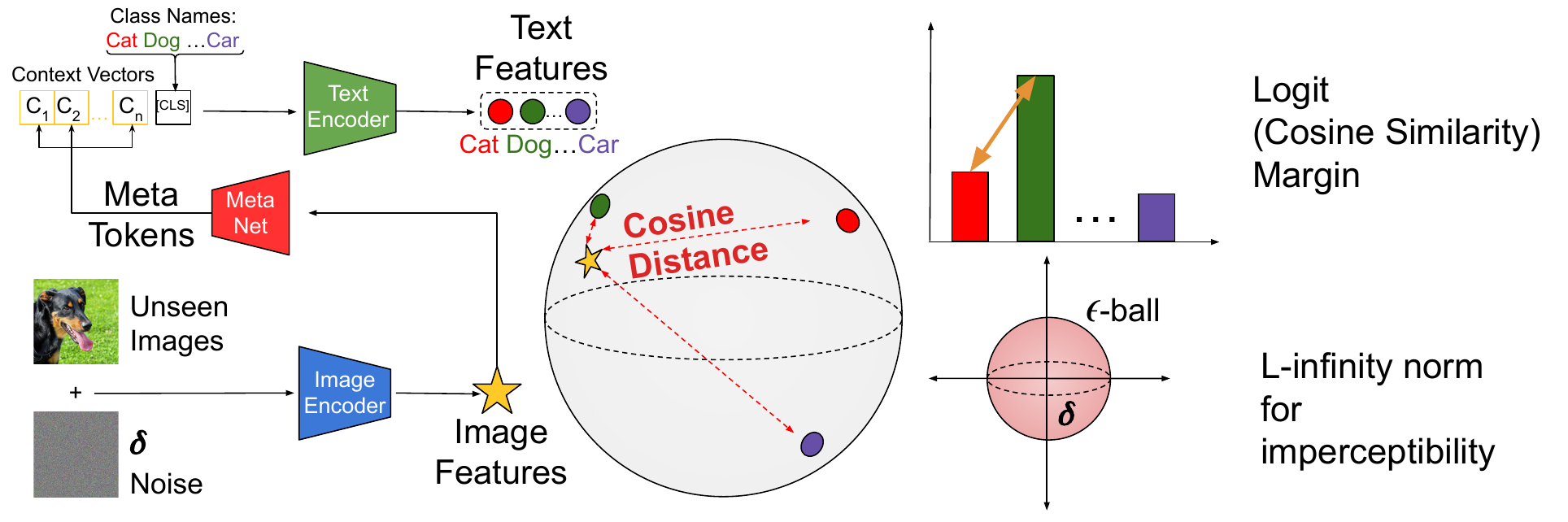}
  \caption{Overview of our CLIP-Inspector framework for detecting backdoor attacks. For each candidate class, CI optimizes an $\ell_\infty$-bounded trigger by minimizing a margin between the target logit and the highest non-target logit on unlabeled OOD images. The triggers are used to compute class-wise anomaly scores with the presence of an outlier yielding a model-level backdoor decision.}
  \label{fig:mainfig}
\end{figure*}

\subsection{Threat Model}
Similar to threats considered in \cite{nguyen2021wanet,bai2024badclip}, our attacker is a model training service provider who has complete control over the prompt-tuning process and maliciously alters it for backdoor purposes. The trained/poisoned model is then delivered to the customer. 
For detection, we assume that the defender is a third-party AI security firm with white-box access to the delivered model. Using our method, they perform gradient-based optimization to reconstruct potential triggers ($\delta_c$) for each class label $c$, without modifying any model parameters.
The defender thus operates under the following constraints:  
\begin{itemize}
  \item White-box access to model architecture and weights.
  \item Small set of unlabeled OOD images ($\sim$1000).
  \item No access to backdoored samples or the backdoor trigger.
  \item A list of potential backdoor target classes.
\end{itemize}

\subsection{Trigger Inversion}
\label{sec:inversion}
Given a prompt-tuned CLIP $(f_I,f_T,h_\theta)$ and a set of unlabeled OOD images $\mathcal{D}$, for each candidate target class $c\in\mathcal{C}$, CI searches for a perturbation $\delta_c$ that forces every unrelated OOD image $x\in\mathcal{D}$ to be classified as $c$. We cast this search as a constrained optimization problem:
\begin{equation}
\begin{aligned}
\min_{\delta}\,&\mathcal{L}_{\text{margin}}(\delta)
 =-\frac{1}{|\mathcal{D}|}
   \sum_{x\in\mathcal{D}}
   \Bigl[l_{c}(x+\delta)-\max_{c'\neq c}l_{c'}(x+\delta)\Bigr] \\
\text{s.t. }
 &\|\delta\|_\infty \le \epsilon
\end{aligned}
\label{eq:constrained_margin}
\end{equation}
where $l_{k}(x)=\text{sim}\bigl(f_I(x),f_T(\{h_\theta(x),k\})\bigr)$
is the cosine similarity for class $k$ and $\epsilon=4/255$ bounds the pixel-level perturbation amplitude to ensure trigger imperceptibility.
Equation \eqref{eq:constrained_margin} defines our \textbf{logit-margin objective}: it lifts the target logit while suppressing only the strongest competitor, focusing gradients on the most relevant class pair and enabling faster convergence to an effective trigger \textit{irrespective of the number of classes}.

We solve the optimization problem via gradient descent. For one epoch (32 optimization steps) over $\mathcal{D}$, the per-step loss for a mini-batch $B$ is given as:
\begin{equation}
\mathcal{L}_{\text{margin}}(\delta)= -\frac{1}{|B|} 
\sum_{x\in B} \Bigl[l_{c}(x+\delta)-\max_{c'\neq c}l_{c'}(x+\delta)\Bigr]
\label{eq:loss}
\end{equation}
After every gradient based update, $\delta$ is projected element‑wise to the $\ell_\infty$ ball $\{\delta:\|\delta\|_\infty\le\epsilon\}$. For the resulting perturbation $\delta_c$ we measure the ASR on a hold-out test-set $\mathcal{T}$ as:
\begin{equation}
  {\rm ASR}(c)=\tfrac{1}{|\mathcal{T}|}\sum_{x\in\mathcal{T}}[\arg\max_{k}\tilde{p}(y = k|x+\delta_c)=c]
\end{equation}
where $\tilde{p}$ is the softmax probability function mentioned previously.
Backdoored classes typically exhibit high ASR while benign classes are near zero.

\subsection{Anomaly Score Calculation}
\label{sec:score}
For each class, we compute an anomaly score based on two key metrics: \textbf{(1) Attack Success Rate (ASR):} Measures how reliably the reconstructed trigger causes misclassification to the target class. \textbf{(2) Average Optimization Loss:} Reflects how easily the optimization converges when targeting a class. A poisoned class exhibits both a \emph{high} ASR and a \emph{low} loss value. We aggregate the $z$-scores for each metric as:
\begin{equation}
S(c)=z_{\text{ASR}}(c)-z_{\text{loss}}(c),
\quad
z_{x}(c)=\frac{x(c)-\mu_x}{\sigma_x},
\label{eq:classwise_anomaly}
\end{equation}

where $\mu_x,\sigma_x$ are the mean and standard deviation over $\mathcal{C}$. 
To obtain a model-level decision without external calibration, we standardize the set $S=\{s_c\}_{c\in \mathcal{C}}$ within the model,
\begin{equation}
\mu_s=\mathrm{mean}(S),\quad \sigma_s=\mathrm{std}(S),\quad
z_c=\frac{s_c-\mu_s}{\sigma_s},
\end{equation}
sort $z_{(1)}\ge z_{(2)}\ge\cdots$, and flag the model as backdoored iff
\begin{equation}
s_{max} = z_{(1)} \;\ge\; k,
\end{equation}
with a default pre-registered constant $k=2$ across datasets. When a small labeled calibration set is available, we suggest replacing the fixed $k$ by a threshold $\tau$ on $s_{\max}$, selected via Youden’s $J$ \cite{Youden1950} on the ROC curve. $\tau$ is then kept fixed for runs with the same $|C|$ and OOD budget.
The end-to-end detection pipeline is illustrated in Figure~\ref{fig:mainfig} and detailed in Algorithm~\ref{alg:ci}.

\begin{algorithm}[btp]
\caption{\textsc{CLIP-Inspector (CI)}}
\label{alg:ci}
\begin{algorithmic}[1]
\STATE \textbf{Input:} Prompt-tuned CLIP $\bigl(f_I,f_T,h_\theta\bigr)$; unlabeled OOD set $\mathcal{D}$;
      candidate labels $\mathcal{C}$; perturbation budget $\epsilon$
\FOR{$c \in \mathcal{C}$}
    \STATE Initialise $\delta \sim \mathcal{U}\bigl[-\epsilon,\epsilon\bigr]$
    \FOR{\textbf{each} mini-batch $B \subset \mathcal{D}$}
        \STATE Compute \\
        $\displaystyle
        \mathcal{L}_{\text{margin}}(\delta)\! =\!
        -\frac{1}{|B|}\sum_{x \in B}
        \Bigl[l_{c}(x{+}\delta)\!-\!\max_{{c'\neq c}} l_{c'}(x{+}\delta)\Bigr]$
        \STATE $\delta \leftarrow \delta - \eta \,\nabla_{\delta}\mathcal{L}_{\text{margin}}$ \hfill \textit{(Adam, $\eta=0.1$)}
        \STATE $\delta \leftarrow \operatorname{clip}\bigl(\delta,-\epsilon,\epsilon\bigr)$
    \ENDFOR
    \STATE Record $\mathcal{L}_{\text{avg}}(c)$ and calculate ${\rm ASR}(c)$
\ENDFOR
\STATE Compute anomaly scores $S=\{s_c\}_{c\in\mathcal{C}}$ (Eq. \ref{eq:classwise_anomaly}), 
\STATE Set $z_c=\dfrac{s_c-\mu_s}{\sigma_s}$, let $z_{(1)}\ge z_{(2)}\ge\cdots$ be the sorted $\{z_c\}$ and $c^\star=\arg\max_{c}z_c$
\STATE \quad \textbf{if} $z_{(1)}\ge k$ \textbf{then} is\_backdoored $\leftarrow$ \texttt{True} \textbf{else} \texttt{False}
\STATE \textbf{return} is\_backdoored, $z_{(1)}$, $c^\star$,$\{\delta_c\}_{c\in\mathcal{C}}$
\end{algorithmic}
\end{algorithm}

\section{Experimental Setup}
\label{sec:exp_setup}

\subsection{Datasets and Models}
We evaluate CLIP-Inspector on ten standard recognition benchmarks: ImageNet~\cite{deng2009imagenet}, Caltech101~\cite{fei2004learning}, OxfordPets~\cite{parkhi2012cats}, Flowers102~\cite{nilsback2008automated}, Food101~\cite{bossard2014food}, FGVC~\cite{maji2013fine}, SUN397~\cite{xiao2010sun}, DTD~\cite{cimpoi2014describing}, EuroSAT~\cite{helber2019eurosat}, and UCF101~\cite{soomro2012ucf101}. 

Unless otherwise stated, we use CLIP ViT-B/16~\cite{radford2021learning} with frozen image and text encoders and prompt-tune it using CoCoOp~\cite{zhou2022conditional}. The meta-net is a two-layer MLP that produces image-conditioned tokens, which are concatenated with four learnable context vectors and class name embeddings to form the prompts (see Figure \ref{fig:mainfig}).

\subsection{Train/Test and OOD Inversion Protocol}
\label{sec:train_test_split}
Following BadCLIP~\cite{bai2024badclip}, we split each dataset into disjoint \emph{Seen} and \emph{Unseen} classes. Prompt-tuning is performed on Seen classes only, and Unseen classes are used for evaluation in a few-shot setting (16 samples per class). The seen class chosen as the backdoor target is added to the unseen set at test time to measure ASR.

For trigger inversion, we sample 50 candidate labels, including the backdoor target, and treat them as the candidate class set $\mathcal{C}$. The inversion pool $D$ consists of 1000 unlabeled images drawn from the Open Images dataset ~\cite{openimages} without class overlap with $\mathcal{C}$, matching our open-set scenario where the attacker aims to misclassify OOD images as in-distribution classes.

\subsection{Backdoor Attacks}
\label{sec:attacks}
We consider four backdoor attacks in the CoCoOp setting. As BadCLIP~\cite{bai2024badclip} was designed for prompt-tuning backdoors, we follow the authors’ setup with a dense $\ell_\infty$-bounded trigger (warmup for 3 epochs and joint optimization of trigger and meta-net for 10 epochs, with $\epsilon = 4/255$).
To test generality beyond a single threat, we adapt three image-space attacks to the prompt-tuning pipeline: Blended~\cite{blended2017}, WaNet~\cite{nguyen2021wanet}, and SIBA~\cite{siba2024backdoor}. We use default hyperparameters and poison 10\% of fine-tuning images for each attack. Overall, these attacks cover dense and sparse, visible and imperceptible, and additive and geometric triggers. More attack-specific details are provided in the supplementary material.

\begin{table}[t]
\centering
\resizebox{\columnwidth}{!}{%
\begin{tabular}{@{}lcccccccc@{}}
\toprule
\multirow{2}{*}{\textbf{Dataset}} &
  \multicolumn{2}{c}{\textbf{\begin{tabular}[c]{@{}c@{}}ACC\\ (Seen Classes)\end{tabular}}} &
  \multicolumn{2}{c}{\textbf{\begin{tabular}[c]{@{}c@{}}ASR\\ (Seen Classes)\end{tabular}}} &
  \multicolumn{2}{c}{\textbf{\begin{tabular}[c]{@{}c@{}}ACC \\ (Unseen Classes)\end{tabular}}} &
  \multicolumn{2}{c}{\textbf{\begin{tabular}[c]{@{}c@{}}ASR \\ (Unseen Classes)\end{tabular}}} \\ \cmidrule(l){2-9} 
        & mean  & std  & mean  & std          & mean  & std  & mean  & std   \\ \midrule
\textbf{Caltech101}   & 97.29\% & 1.51\% & 89.73\% & \multicolumn{1}{c|}{17.59\%} & 92.25\% & 1.28\% & 79.89\% & 28.13\% \\
\textbf{DTD}    & 71.10\% & 6.48\% & 88.92\% & \multicolumn{1}{c|}{9.08\%}  & 44.24\% & 3.67\% & 83.25\% & 11.49\% \\
\textbf{EuroSAT}    & 89.45\% & 5.40\% & 95.35\% & \multicolumn{1}{c|}{6.23\%}  & 46.25\% & 2.79\% & 76.25\% & 26.79\% \\
\textbf{FGVC} & 36.06\% & 3.67\% & 89.01\% & \multicolumn{1}{c|}{20.78\%} & 30.89\% & 2.02\% & 69.76\% & 38.80\% \\
\textbf{Food101}    & 89.23\% & 0.60\% & 93.64\% & \multicolumn{1}{c|}{10.46\%} & 89.04\% & 1.89\% & 91.83\% & 12.17\% \\
\textbf{ImageNet}   & 74.19\% & 1.96\% & 78.84\% & \multicolumn{1}{c|}{39.32\%} & 67.13\% & 1.75\% & 75.31\% & 41.10\% \\
\textbf{Flowers102}   & 93.55\% & 3.12\% & 97.82\% & \multicolumn{1}{c|}{2.89\%}  & 71.65\% & 1.24\% & 97.64\% & 3.75\%  \\
\textbf{OxfordPets}   & 92.16\% & 3.48\% & 81.99\% & \multicolumn{1}{c|}{24.01\%} & 92.18\% & 4.99\% & 73.36\% & 24.64\% \\
\textbf{SUN397}   & 76.93\% & 3.27\% & 81.73\% & \multicolumn{1}{c|}{34.19\%} & 72.49\% & 2.97\% & 80.70\% & 34.16\% \\
\textbf{UCF101}   & 83.54\% & 1.52\% & 96.94\% & \multicolumn{1}{c|}{3.29\%}  & 69.17\% & 2.93\% & 92.46\% & 8.20\%  \\ \bottomrule
\end{tabular}%
}
\caption{Accuracy and Attack success rate for all datasets averaged across 4 attack types and clean models. }
\label{tab:averagedataset}
\end{table}


\begin{figure*}[t]
  \centering
  \begin{subfigure}[b]{0.3\linewidth}
    \centering
    \includegraphics[width=\linewidth]{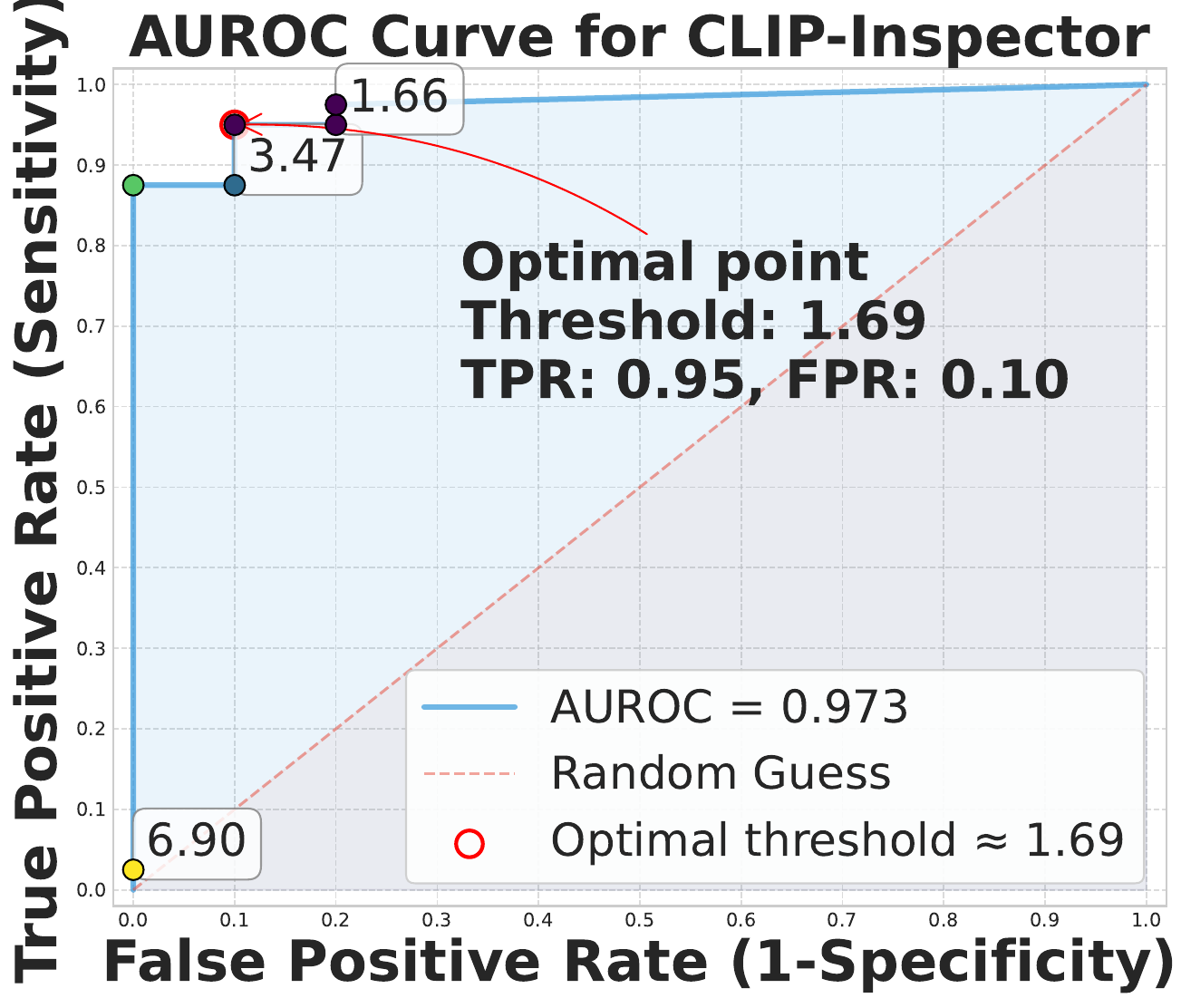}
    \caption{CI}
    \label{fig:CI_AUROC}
  \end{subfigure}\hfill
  \begin{subfigure}[b]{0.3\linewidth}
    \centering
    \includegraphics[width=\linewidth]{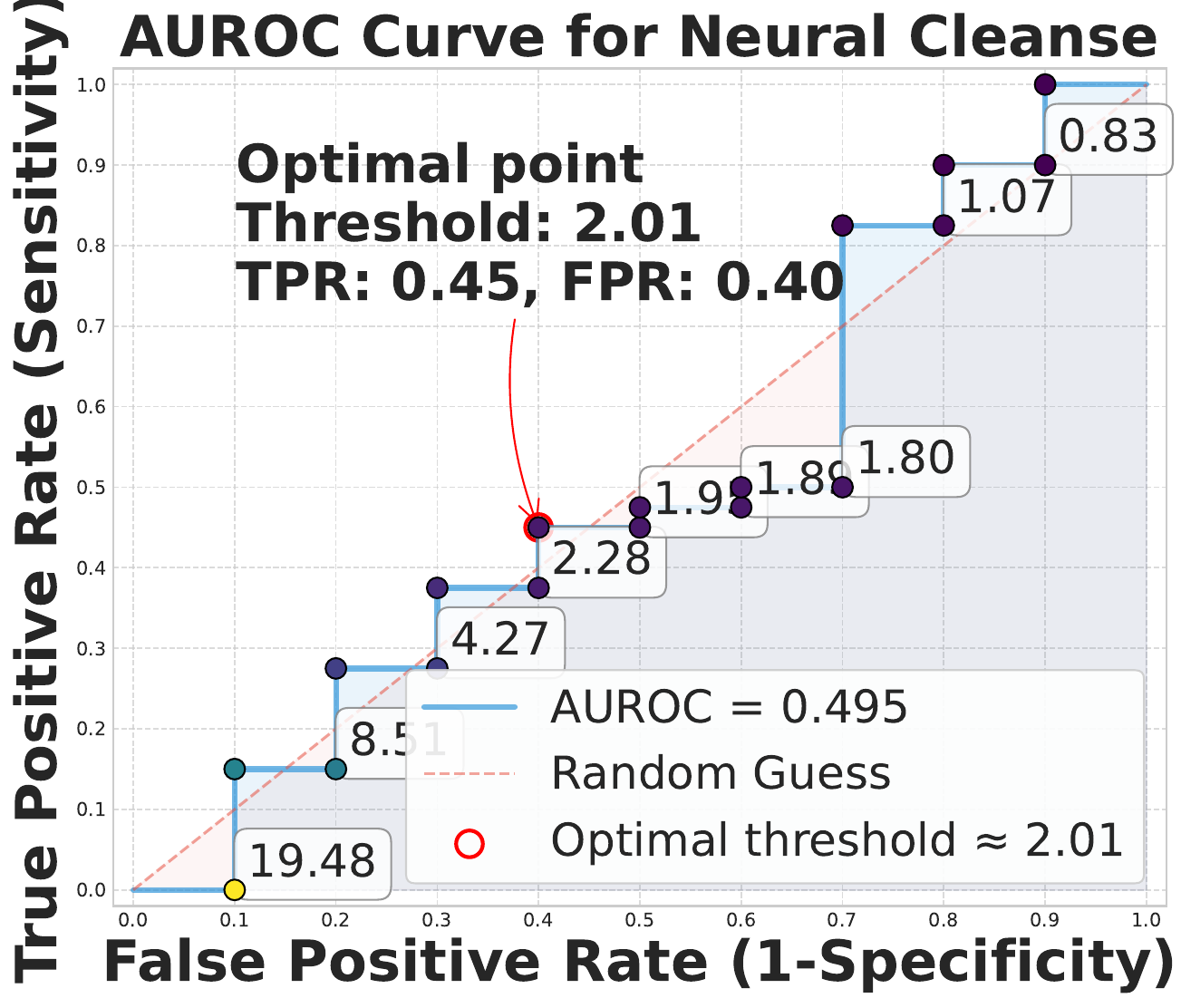}
    \caption{NC}
    \label{fig:NC_AUROC}
  \end{subfigure}\hfill
  \begin{subfigure}[b]{0.3\linewidth}
    \centering
    \includegraphics[width=\linewidth]{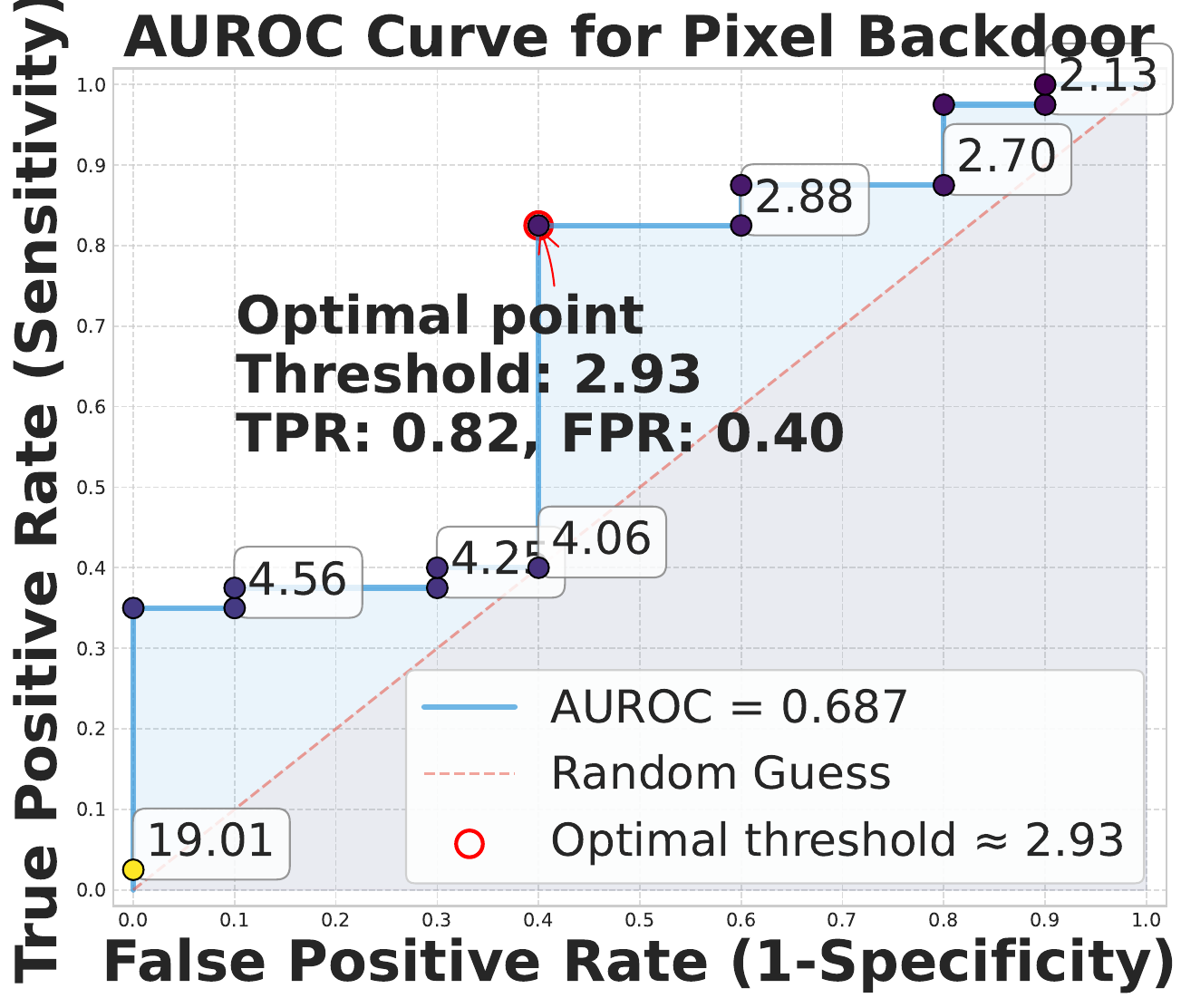}
    \caption{PixB}
    \label{fig:pixb_AUROC}
  \end{subfigure}
  \vspace{0.1cm}
  \begin{subfigure}[b]{0.7\linewidth}
    \centering
    \includegraphics[width=1\linewidth]{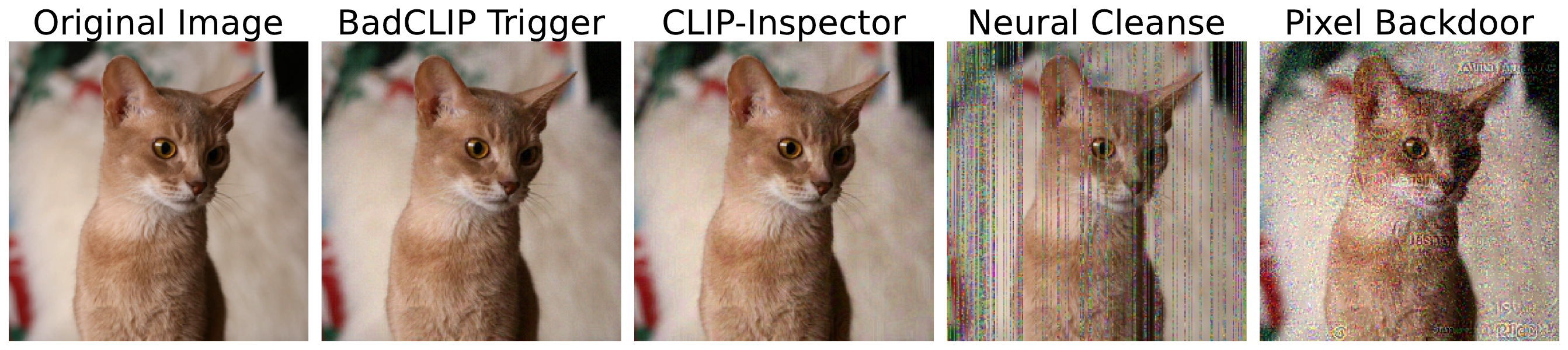}
    \caption{From left to right, Original image without trigger, Image with BadCLIP trigger, Image with triggers reconstructed via CI, NC and PixB respectively.}
    \label{fig:recon_images}
  \end{subfigure}
  \caption{AUROC plots for each detection method. Our method achieves a high AUROC of 0.973, whereas NC and PixB have low scores of 0.495 and 0.687, respectively. Our method also shows a clear distinction between clean and backdoored models in terms of average anomaly score.   }
  \label{fig:results_main}
\end{figure*}

\subsection{Baselines}
\label{sec:baselines}
We adapt Neural Cleanse (NC)~\cite{wang2019neural} and Pixel Backdoor (PixB)~\cite{pixelbackdoor2022} to the prompt-tuning setting. None of the other methods is compatible, as they either analyze encoder outputs or operate at test time. NC and PixB both optimize class-wise triggers and flag anomalous classes based on trigger properties, with PixB being the current state-of-the-art. NC learns a masked patch and utilizes the $\ell_1$ norm of the mask, whereas PixB optimizes per-pixel positive and negative perturbations and uses the $\ell_0$ norm. For both methods, we keep their original loss functions and dynamic regularization techniques. Note that we do not modify model architecture in any case.

\subsection{Metrics and Implementation Details}
\label{sec:metrics_impl}
At the \emph{model} level, we report AUROC over the max anomaly score ($s_{\max}$) to measure how well a method separates clean from backdoored models and provide a calibrated decision threshold. At the \emph{class} level, we report F1-score over $\mathcal{C}$ to quantify how accurately each detector identifies the backdoor target class.

All methods are implemented in PyTorch 2.1 with Python 3.10 and run on an Ubuntu 20.04 machine with a 32-core CPU and one A100 80GB GPU. For CI, we use Adam with a step size ($\eta$) of 0.1, a batch size of 32, 1000 OOD images, and a single epoch of optimization per class ($\approx32$ gradient steps). On the same dataset, NC and PixB are run for 5 epochs with dynamic regularization. Note that a single epoch is sufficient for CI, while other methods need multiple iterations.

\section{Results and Analysis}
\label{sec:results}

\subsection{Overall Detection Performance}
\label{sec:overall_perf}

For each dataset, we train one clean and four backdoored, prompt-tuned CLIP models (one per attack), yielding a total of 50 models. CI, NC, and PixB are run on every model using the same candidate class set $\mathcal{C}$ and OOD inversion pool. Each method outputs class-wise anomaly scores, and a model is marked backdoored if at least one class exceeds the method-specific anomaly threshold.

We fix the operating point at k=2 for each detector and also plot the ROC curve over model-level anomaly scores (Figure~\ref{fig:results_main}). CI achieves a detection accuracy of 94\% (47/50 models) with AUROC 0.973, while NC and PixB obtain AUROCs of 0.495 and 0.687, respectively. At the class level, CI reaches an average F1 of 0.92 across datasets (Table~\ref{tab:fpdataset}), with almost zero false positives. NC fails to reliably identify the target class (average F1 0.06) and generates many spurious anomalies, whereas PixB performs moderately (average F1 0.42) but still mislabels many benign classes. Beyond aggregate AUROC, we report per-attack/per-dataset statistics and threshold sensitivity in the supplementary material.

\begin{table}[t]
\centering
\resizebox{\columnwidth}{!}{%
\begin{tabular}{@{}lccccccccccccc@{}}
\toprule
 & \multicolumn{1}{l}{} & \multicolumn{4}{c}{\textbf{CI (Ours)}} & \multicolumn{4}{c}{\textbf{NC}} & \multicolumn{4}{c}{\textbf{PixB}} \\ \cmidrule(l){2-14} 
\textbf{Dataset} & \begin{tabular}[c]{@{}c@{}}Total \\ Classes\end{tabular} & TP & FP & FN & F1 & TP & FP & FN & F1 & TP & FP & FN & F1 \\ \midrule
\multicolumn{1}{l|}{\textbf{Caltech101}} & \multicolumn{1}{c|}{250} & 4 & 0 & 0 & \multicolumn{1}{c|}{\textbf{1.00}} & 1 & 0 & 3 & \multicolumn{1}{c|}{0.40} & 2 & 6 & 2 & 0.33 \\
\multicolumn{1}{l|}{\textbf{DTD}} & \multicolumn{1}{c|}{235} & 3 & 0 & 1 & \multicolumn{1}{c|}{\textbf{0.86}} & 0 & 13 & 4 & \multicolumn{1}{c|}{0.00} & 1 & 1 & 3 & 0.33 \\
\multicolumn{1}{l|}{\textbf{EuroSAT}} & \multicolumn{1}{c|}{50} & 2 & 1 & 2 & \multicolumn{1}{c|}{\textbf{0.57}} & 0 & 1 & 4 & \multicolumn{1}{c|}{0.00} & 2 & 2 & 2 & 0.50 \\
\multicolumn{1}{l|}{\textbf{FGVC}} & \multicolumn{1}{c|}{250} & 4 & 2 & 0 & \multicolumn{1}{c|}{\textbf{0.80}} & 0 & 21 & 4 & \multicolumn{1}{c|}{0.00} & 2 & 2 & 2 & 0.50 \\
\multicolumn{1}{l|}{\textbf{Food101}} & \multicolumn{1}{c|}{250} & 4 & 0 & 0 & \multicolumn{1}{c|}{\textbf{1.00}} & 2 & 21 & 2 & \multicolumn{1}{c|}{0.15} & 3 & 8 & 1 & 0.40 \\
\multicolumn{1}{l|}{\textbf{ImageNet}} & \multicolumn{1}{c|}{250} & 4 & 0 & 0 & \multicolumn{1}{c|}{\textbf{1.00}} & 1 & 22 & 3 & \multicolumn{1}{c|}{0.07} & 3 & 2 & 1 & 0.67 \\
\multicolumn{1}{l|}{\textbf{Flowers102}} & \multicolumn{1}{c|}{250} & 4 & 0 & 0 & \multicolumn{1}{c|}{\textbf{1.00}} & 0 & 37 & 4 & \multicolumn{1}{c|}{0.00} & 3 & 7 & 1 & 0.43 \\
\multicolumn{1}{l|}{\textbf{OxfordPets}} & \multicolumn{1}{c|}{185} & 4 & 0 & 0 & \multicolumn{1}{c|}{\textbf{1.00}} & 0 & 2 & 4 & \multicolumn{1}{c|}{0.00} & 2 & 9 & 2 & 0.27 \\
\multicolumn{1}{l|}{\textbf{SUN397}} & \multicolumn{1}{c|}{250} & 4 & 0 & 0 & \multicolumn{1}{c|}{\textbf{1.00}} & 0 & 18 & 4 & \multicolumn{1}{c|}{0.00} & 4 & 10 & 0 & 0.44 \\
\multicolumn{1}{l|}{\textbf{UCF101}} & \multicolumn{1}{c|}{250} & 4 & 0 & 0 & \multicolumn{1}{c|}{\textbf{1.00}} & 0 & 5 & 4 & \multicolumn{1}{c|}{0.00} & 2 & 8 & 2 & 0.29 \\ \midrule
\multicolumn{1}{l|}{\textbf{Average}} & \multicolumn{1}{c|}{222} & 3.7 & 0.3 & 0.3 & \multicolumn{1}{c|}{\textbf{0.92}} & 0.4 & 14 & 3.6 & \multicolumn{1}{c|}{0.06} & 2.4 & 5.5 & 1.6 & 0.42 \\\bottomrule
\end{tabular}%
}
\caption{Target Class detection metrics for each method. CLIP Inspector has almost zero false positives, correctly identifying the target class for every dataset.}
\label{tab:fpdataset}
\end{table}

\begin{table}[t]
\centering
\resizebox{\columnwidth}{!}{%
\begin{tabular}{@{}lccccccccccccc@{}}
\toprule
\multirow{2}{*}{\textbf{Attack}} & \multicolumn{1}{l}{\textbf{}} & \multicolumn{4}{c}{\textbf{CI (Ours)}} & \multicolumn{4}{c}{\textbf{NC}} & \multicolumn{4}{c}{\textbf{PixB}} \\ \cmidrule(l){2-14} 
 & \begin{tabular}[c]{@{}c@{}}Total \\ Classes\end{tabular} & TP & FP & FN & F1 & TP & FP & FN & F1 & TP & FP & FN & F1 \\ \midrule
\multicolumn{1}{l|}{\textbf{BadCLIP}} & \multicolumn{1}{c|}{444} & 9 & 0 & 1 & \multicolumn{1}{c|}{\textbf{0.94}} & 1 & 30 & 9 & \multicolumn{1}{c|}{0.04} & 6 & 11 & 4 & 0.44 \\
\multicolumn{1}{l|}{\textbf{Blended}} & \multicolumn{1}{c|}{444} & 10 & 1 & 0 & \multicolumn{1}{c|}{\textbf{0.95}} & 2 & 24 & 8 & \multicolumn{1}{c|}{0.11} & 10 & 8 & 0 & 0.71 \\
\multicolumn{1}{l|}{\textbf{SIBA}} & \multicolumn{1}{c|}{444} & 10 & 0 & 0 & \multicolumn{1}{c|}{\textbf{1}} & 0 & 32 & 10 & \multicolumn{1}{c|}{0} & 6 & 15 & 4 & 0.38 \\
\multicolumn{1}{l|}{\textbf{WaNet}} & \multicolumn{1}{c|}{444} & 8 & 1 & 2 & \multicolumn{1}{c|}{\textbf{0.84}} & 1 & 35 & 9 & \multicolumn{1}{c|}{0.04} & 2 & 11 & 8 & 0.17 \\ \midrule
\textbf{Average} & 444 & 9.25 & 0.5 & 0.75 & \textbf{0.93} & 1 & 30.25 & 9 & 0.05 & 6 & 11.25 & 4 & 0.42 \\ \bottomrule
\end{tabular}%
}
\caption{Target class detection metrics for each attack. CI has a high F1 score across all attack types.}
\label{tab:fpattacks}
\end{table}

\subsubsection{Loss function choice}
\label{sec:loss_ablation}

CI optimizes a single perturbation $\delta_c$ per class by minimizing a margin-based objective $L_{\text{margin}}$ over a set of unlabeled OOD images. To assess whether this choice is crucial, we compare it against two natural alternatives: cross-entropy on the target class and raw logit maximization.

Given logits $\l_k(x)$ for class $k$ and input $x$, and an OOD inversion set $D$, we consider:
\begin{align}
\mathcal{L}_{\text{logit}}(\delta) &= - \frac{1}{|D|} \sum_{x \in D} \l_c(x+\delta),\\
\mathcal{L}_{\text{CE}}(\delta) &= \frac{1}{|D|} \sum_{x \in D} 
\text{CE}\bigl( \text{softmax}(\l(x+\delta)), c \bigr), \\
\mathcal{L}_{\text{margin}}(\delta) &= - \frac{1}{|D|} \sum_{x \in D} 
\bigl( l_c(x+\delta) - \max_{k \neq c} \l_k(x+\delta) \bigr),
\end{align}
where $c$ is the candidate target class. For each loss, we run CI on a representative attack (BadCLIP) and evaluate the resulting perturbations' \emph{attack success rate}. With identical budgets/pools, $L_{\text{margin}}$ reaches 91.09\% backdoor-class ASR vs 81.45\% ($L_{\text{CE}}$) and 77.66\% ($L_{\text{logit}}$), while keeping benign-class ASR lowest (4.62\%), yielding a larger clean-vs-poisoned separation (Table 4).
These results empirically support our choice of a margin-based objective: by explicitly enforcing a gap between the target and the strongest non-target class, $L_{\text{margin}}$ produces more backdoor-like perturbations and more discriminative anomaly scores. We justify other design choices like the usage of OOD vs ID data, batch size, and number of OOD samples via ablations in the supplementary material.

\begin{table}[]
\centering
\small
\resizebox{\columnwidth}{!}{%
\begin{tabular}{@{}lcccc@{}}
\toprule
\multirow{2}{*}{\textbf{Loss Type}} & \multicolumn{2}{c}{\textbf{Backdoor class ASR}} & \multicolumn{2}{c}{\textbf{Average other class ASR}} \\ \cmidrule(l){2-5} 
                            & Mean  & Std                        & Mean & Std  \\ \midrule
\multicolumn{1}{l|}{$\mathcal{L}_{\text{logit}}$}  & 77.66 & \multicolumn{1}{c|}{26.96} & 5.37 & 3.11 \\
\multicolumn{1}{l|}{$\mathcal{L}_{\text{CE}}$}     & 81.45 & \multicolumn{1}{c|}{13.48} & 5.26 & 0.83 \\
\multicolumn{1}{l|}{$\mathcal{L}_{\text{margin}}$} & \textbf{91.09} & \multicolumn{1}{c|}{8.73}  & \textbf{4.62} & 0.8  \\ \bottomrule
\end{tabular}%
}
\caption{Comparison of different loss functions in terms of reconstructed trigger ASR, averaged over 4 datasets (ImageNet, Caltech101, UCF101, DTD) for BadCLIP (ASR$\approx99\%$).}
\label{tab:loss_ablation}
\end{table}

\subsection{Adaptive Attack}
We next evaluate CLIP‑Inspector’s robustness against an \emph{adaptive} adversary who is aware of our detection strategy.  
Recall that standard BadCLIP learns a universal additive trigger $\delta$ that is applied to every training image, constrained by $\|\delta\|_{\infty} \leq \epsilon$ (with $\delta_{init} \sim \mathrm{Uniform}(-\epsilon, \epsilon)/\sigma$), where $\sigma$ is CLIP's per channel RGB standard deviation vector used for normalization.
Likewise, CLIP‑Inspector reconstructs candidate triggers for a frozen prompt‑tuned model by initialising and optimizing $\delta_c$ for each class $c \in \mathcal{C}$.  
An attacker could attempt to undermine this process by enforcing \emph{specificity},
ensuring that the backdoor activates only for the exact trigger pattern and not for any of its close approximations, which is what CLIP‑Inspector aims to recover.

To model this threat, we propose \textbf{BadCLIP Adaptive}, a two‑phase variant of BadCLIP designed to reduce the ASR of reconstructed triggers.  
\textbf{Phase 1} mirrors standard BadCLIP and produces an optimal trigger $\delta^{\star}$, which is then frozen as $\delta_{\text{fixed}}$.  
\textbf{Phase 2} further trains the model to be sensitive to perturbations by generating
$\delta_{\text{perturbed}} = \delta_{\text{fixed}} + \eta$,
where $\eta \sim \mathcal{N}(0, \alpha*\epsilon/\sigma)$
and $\alpha$ controls the perturbation strength; the resulting
$\delta_{\text{perturbed}}$ is clipped to $[-\epsilon/\sigma, \epsilon/\sigma]$.  
Therefore, Phase 2 uses three image types: clean images $X$ (label $y$),
exact‑trigger images $X + \delta_{\text{fixed}}$ (label $y_t$),
and perturbed‑trigger images $X + \delta_{\text{perturbed}}$ (label $y$).  
This forces the model to associate only the \emph{exact} trigger with the target class,
creating a narrow “basin of attraction’’ that is likely to frustrate CLIP‑Inspector’s trigger inversion process.
Empirically, we observe that BadCLIP Adaptive achieves a substantially lower ASR than standard BadCLIP, decreasing by 30-40\% on average. The ASR of CLIP‑Inspector’s reconstructed triggers also declines by around 20\% on average. However, it is still able to reliably detect backdoors, showcasing our method's robustness to adaptive attacks. More details in supplementary material.

\subsection{Removal using reconstructed trigger}
So far, we have used CI's reconstructed perturbations purely as a diagnostic tool for model-level detection. Since these perturbations are optimized to mimic the effect of a potential backdoor trigger for class $c$, a natural question is if they can also be used for \emph{post-hoc backdoor removal}.

In this subsection, we consider a defender who, in addition to white-box access to the delivered model, also has access to a small labeled clean dataset for the downstream task (e.g., a subset of the data used for prompt tuning). We first run CI as usual to reconstruct $\delta_c$ for each class and identify the class $\hat{c}$ with the highest anomaly score $s_{max}$. We then treat $\delta_{\hat{c}}$ as a surrogate for the unknown backdoor trigger and run CoCoOp-style fine-tuning on triggered inputs $(x + \delta_{\hat{c}}, y)$, where $x$ is drawn from the clean dataset and $y$ is its ground-truth label. The objective is to re-align the model so that it no longer misclassifies triggered inputs into $\hat{c}$.

We evaluate this strategy on the same backdoored models used in our detection experiments. Table \ref{tab:backdoor_removal} reports the backdoor attack success rate (ASR) on OOD test images and the clean top-1 accuracy (ACC) before and after CI-based fine-tuning. For all attack types, a small number of gradient steps on triggered inputs is enough to reduce the ASR to $<10\%$ while preserving clean accuracy ($\pm 1\%$ of the original model). This indicates that CI's reconstructed perturbations are not mere artifacts, but are functionally close to the attacker’s trigger and can be utilized for backdoor removal.
We emphasize that detection itself still requires only unlabeled OOD images, and labels are used only for this optional repair step.

\subsection{Generalizability beyond prompt tuning}
\label{sec:gen_beyond_prompt}
CI is agnostic to where the backdoor resides as it probes a delivered classifier via OOD trigger inversion and behaviour-based scoring, without assuming a meta-network. To validate this, we poison the image-encoder by fully fine-tuning CLIP with Blended-style triggers~\cite{blended2017} (no prompt/meta-net), using three patterns (random-noise patch, triangle, text). We then run CI \emph{unchanged} (same $|C|$, OOD pool, $\ell_\infty$ budget). In all cases, CI reconstructs perturbations with high ASR on the true backdoor class, correctly identifying clean vs.\ poisoned models. More details in the supplementary.

\begin{table}[t]
\centering
\resizebox{\columnwidth}{!}{%
\begin{tabular}{@{}lcccc@{}}
\toprule
\textbf{Attack}                       & \textbf{ACC before} & \textbf{ACC after}                    & \textbf{ASR before} & \textbf{ASR after} \\ \midrule
\multicolumn{1}{l|}{\textbf{BadCLIP}}           & 81.76\%    & \multicolumn{1}{c|}{81.35\%} & 98.48\%    & 7.61\%    \\
\multicolumn{1}{l|}{\textbf{BadCLIP\_adaptive}} & 81.62\%    & \multicolumn{1}{c|}{81.04\%} & 64.35\%    & 4.23\% \\
\multicolumn{1}{l|}{\textbf{Blended}}           & 81.69\%    & \multicolumn{1}{c|}{81.69\%} & 99.73\%    & 7.64\%    \\
\multicolumn{1}{l|}{\textbf{SIBA}}              & 75.91\%    & \multicolumn{1}{c|}{77.93\%} & 66.30\%    & 5.76\%    \\
\multicolumn{1}{l|}{\textbf{WaNet}}             & 82.08\%    & \multicolumn{1}{c|}{82.11\%} & 92.99\%    & 5.27\%    \\ \midrule
\multicolumn{1}{l|}{\textbf{Average}}           & 80.61\%    & \multicolumn{1}{c|}{80.82\%} & 84.37\%    & 6.10\%    \\ \bottomrule
\end{tabular}%
}
\caption{ACC and ASR for all attack types averaged over 10 datasets. Fine-tuning using reconstructed triggers effectively reduces the backdoor ASR to below 10\%.}
\label{tab:backdoor_removal}
\end{table}
\section{Discussion}\label{sec:discuss}
Our detector succeeds because \textit{poisoned} classes simultaneously exhibit \textbf{high} ASR and \textbf{low} optimization loss, yielding anomaly scores that stand out as clear outliers. The AUROC curve stays flat for thresholds $>2$ in our experiments, so the default k=2 works well. In practice, we suggest threshold calibration using a small set of clean and poisoned models.  
Failure arises on three (dataset, attack) pairs—\textit{(FGVC, clean)}, \textit{(DTD, WaNet)}, and \textit{(EuroSAT, BadCLIP)}. These failures occur on fine-grained datasets with low inter-class separability, evident by the low Unseen accuracy in Table \ref{tab:averagedataset}. This suggests CI's anomaly gap shrinks when the base classification task is intrinsically ambiguous, revealing a key limitation.

CI needs only \textbf{1000} unlabeled images, and more data doesn't lead to any significant gain in accuracy. Cutting to 500 lowers reconstructed-trigger ASR by 5–10 \%, and cutting to 100 lowers it by 50–60 \%.  Thus, we recommend at least 1000 images for robust inversion. Similarly, running CI for more than 1 epoch doesn't lead to significant changes in performance. The optimization is stable and quickly converges to a minimum in the presence of a backdoor. CI completes a 50-class scan in \textless1h on an A100 GPU. Neural Cleanse and Pixel-Backdoor require 6h and 8h, respectively, and still underperform, thus highlighting CI’s practicality for rapid pre-deployment checks.

The inverted triggers remain \emph{imperceptible}: BadCLIP’s ground-truth Structural Similarity score or SSIM is 0.96. CI's reconstructed triggers have an average SSIM of 0.93, while Neural Cleanse and Pixel-Backdoor degrade to 0.76 and 0.63, respectively (Figure \ref{fig:recon_images}). Moreover, the inverted triggers are functionally similar to the original attacker, as evidenced by their ability to significantly reduce ASR through fine-tuning.
Additionally, since we make no assumptions regarding the presence of a meta-net, our method generalizes well beyond prompt-tuned backdoors, as demonstrated by its ability to reconstruct blended triggers embedded into CLIP's image encoder using the same mechanism.
\section{Limitations}
Despite its effectiveness, our approach has several limitations that suggest directions for future research:

\textbf{Trigger Types:} Our method is able to detect sparse and dense, visible and imperceptible triggers. However, its performance remains unverified on semantic triggers. \textit{Such attacks have not yet been demonstrated on CLIP models}. 

\textbf{Reliance on Dataset Separability:} As already discussed, our approach relies on the model's classification capability. However, we argue that such a model would be low-performing and thus would not be worth backdooring.

\section{Conclusion}
We present \textbf{CLIP‑Inspector}, a lightweight model vetting method to uncover prompt‑tuned backdoors in CLIP models before deployment. By optimizing a dense $\ell_\infty$‑bounded perturbation on unlabeled OOD images and scoring classes via behavioural signals (ASR+ margin-loss), CI detects 94\% of poisoned models in a single epoch over 1000 images. We also clarify how backdoors work in the prompt-tuning setting: a tiny change in the image embedding is amplified by the meta-network, leading to significant shifts in text embeddings. Our method is efficient and practical, and performs significantly better than adapted state-of-the-art methods. Increasing the performance of our method for fine-grained datasets, where class boundaries blur, and reducing the number of false positives to zero remains a challenge that we plan to address in future work.

\section{Ethical Statement}
Our work aims to reduce real-world harm by enabling pre-deployment detection of prompt-tuned backdoors. To prevent adversaries from learning about our defense, our code release would be done in a staged manner, initially to security researchers, and only made public once mitigations are in place. False positives can be minimized by calibrating the detection threshold, and human review is recommended for a suspected model.
{
    \small
    \bibliographystyle{ieeenat_fullname}
    \bibliography{citations}
}

\clearpage
\maketitlesupplementary

\section{ACC and ASR for all attack types}
\subsection{Clean model training}
We prompt-tuned the CLIP model with CoCoOp on 10 image-classification datasets. Accuracy (ACC) values for the seen and unseen subsets of classes for each dataset are shown in Table \ref{tab:acc_asr_stats_clean}. DTD, EuroSAT, and FGVC datasets achieve the lowest cross-domain (unseen) accuracy, highlighting that prompt tuning is less effective for them.

\begin{table}[H]
\centering
\resizebox{\columnwidth}{!}{%
\begin{tabular}{@{}lcccccc@{}}
\toprule
\textbf{Dataset} & \textbf{\makecell{Seen \\ ACC}} & \textbf{\makecell{Seen \\ ASR}} & \textbf{\makecell{Unseen \\ ACC}} & \textbf{\makecell{Unseen \\ ASR}} & \textbf{\makecell{ACC \\ Diff.}} & \textbf{\makecell{ASR \\ Diff.}} \\ \midrule
\multicolumn{1}{l|}{\textbf{Caltech101}} & 98.19\% & \multicolumn{1}{c|}{-} & 93.12\% & \multicolumn{1}{c|}{-} & 5.07\% & - \\
\multicolumn{1}{l|}{\textbf{DTD}} & 75.60\% & \multicolumn{1}{c|}{-} & 50.46\% & \multicolumn{1}{c|}{-} & 25.14\% & - \\
\multicolumn{1}{l|}{\textbf{EuroSAT}} & 91.14\% & \multicolumn{1}{c|}{-} & 47.54\% & \multicolumn{1}{c|}{-} & 43.60\% & - \\
\multicolumn{1}{l|}{\textbf{FGVC\_Aircraft}} & 37.15\% & \multicolumn{1}{c|}{-} & 31.97\% & \multicolumn{1}{c|}{-} & 5.18\% & - \\
\multicolumn{1}{l|}{\textbf{Food101}} & 90.26\% & \multicolumn{1}{c|}{-} & 90.93\% & \multicolumn{1}{c|}{-} & -0.67\% & - \\
\multicolumn{1}{l|}{\textbf{ImageNet}} & 76.05\% & \multicolumn{1}{c|}{-} & 69.39\% & \multicolumn{1}{c|}{-} & 6.66\% & - \\
\multicolumn{1}{l|}{\textbf{Flowers102}} & 94.30\% & \multicolumn{1}{c|}{-} & 73.62\% & \multicolumn{1}{c|}{-} & 20.68\% & - \\
\multicolumn{1}{l|}{\textbf{OxfordPets}} & 95.68\% & \multicolumn{1}{c|}{-} & 96.66\% & \multicolumn{1}{c|}{-} & -0.98\% & - \\
\multicolumn{1}{l|}{\textbf{SUN397}} & 79.46\% & \multicolumn{1}{c|}{-} & 75.54\% & \multicolumn{1}{c|}{-} & 3.92\% & - \\
\multicolumn{1}{l|}{\textbf{UCF101}} & 84.06\% & \multicolumn{1}{c|}{-} & 73.62\% & \multicolumn{1}{c|}{-} & 10.44\% & - \\ \midrule
\textbf{Average} & \textbf{82.19\%} & \textbf{-} & \textbf{70.29\%} & \textbf{-} & \textbf{11.90\%} & \textbf{-} \\ \bottomrule
\end{tabular}%
}
\caption{Accuracy of clean models on Seen and Unseen class subsets for each dataset.}
\label{tab:acc_asr_stats_clean}
\end{table}

\subsection{BadCLIP attack performance}
The BadCLIP attack produces imperceptible triggers ($\ell_\infty\leq4/255$) that transfer well across domains, evident by the high Attack success rate (ASR) values ($>90\%$) on unseen classes for all datasets.

\begin{table}[H]
\centering
\resizebox{\columnwidth}{!}{%
\begin{tabular}{@{}lcccccc@{}}
\toprule
\textbf{Dataset} & \textbf{\makecell{Seen \\ ACC}} & \textbf{\makecell{Seen \\ ASR}} & \textbf{\makecell{Unseen \\ ACC}} & \textbf{\makecell{Unseen \\ ASR}} & \textbf{\makecell{ACC \\ Diff.}} & \textbf{\makecell{ASR \\ Diff.}} \\ \midrule
\multicolumn{1}{l|}{\textbf{Caltech101}} & 98.06\% & \multicolumn{1}{c|}{99.81\%} & 93.78\% & \multicolumn{1}{c|}{99.56\%} & 4.28\% & 0.25\% \\
\multicolumn{1}{l|}{\textbf{DTD}} & 74.03\% & \multicolumn{1}{c|}{94.20\%} & 43.87\% & \multicolumn{1}{c|}{90.16\%} & 30.16\% & 4.04\% \\
\multicolumn{1}{l|}{\textbf{EuroSAT}} & 91.40\% & \multicolumn{1}{c|}{99.95\%} & 47.85\% & \multicolumn{1}{c|}{96.72\%} & 43.55\% & 3.23\% \\
\multicolumn{1}{l|}{\textbf{FGVC\_Aircraft}} & 36.19\% & \multicolumn{1}{c|}{99.94\%} & 31.37\% & \multicolumn{1}{c|}{91.72\%} & 4.82\% & 8.22\% \\
\multicolumn{1}{l|}{\textbf{Food101}} & 90.09\% & \multicolumn{1}{c|}{99.71\%} & 90.89\% & \multicolumn{1}{c|}{98.48\%} & -0.80\% & 1.23\% \\
\multicolumn{1}{l|}{\textbf{ImageNet}} & 75.72\% & \multicolumn{1}{c|}{99.60\%} & 69.72\% & \multicolumn{1}{c|}{99.02\%} & 6.00\% & 0.58\% \\
\multicolumn{1}{l|}{\textbf{Flowers102}} & 93.83\% & \multicolumn{1}{c|}{99.91\%} & 72.20\% & \multicolumn{1}{c|}{100.00\%} & 21.63\% & -0.09\% \\
\multicolumn{1}{l|}{\textbf{OxfordPets}} & 94.50\% & \multicolumn{1}{c|}{98.09\%} & 92.74\% & \multicolumn{1}{c|}{94.01\%} & 1.76\% & 4.08\% \\
\multicolumn{1}{l|}{\textbf{SUN397}} & 79.00\% & \multicolumn{1}{c|}{99.59\%} & 76.86\% & \multicolumn{1}{c|}{98.37\%} & 2.14\% & 1.22\% \\
\multicolumn{1}{l|}{\textbf{UCF101}} & 84.49\% & \multicolumn{1}{c|}{99.63\%} & 69.49\% & \multicolumn{1}{c|}{98.99\%} & 15.00\% & 0.64\% \\ \midrule
\textbf{Average} & \textbf{81.73\%} & \textbf{99.04\%} & \textbf{68.88\%} & \textbf{96.70\%} & \textbf{12.85\%} & \textbf{2.34\%} \\ \bottomrule
\end{tabular}%
}
\caption{ACC and ASR of BadCLIP-poisoned models on Seen and Unseen subsets.}
\label{tab:acc_asr_stats_badclip}
\end{table}

\subsection{Blended attack performance}
The Blended attack performs on par with BadCLIP, achieving strong levels of cross-domain transfer despite a static trigger. However, the trigger is not imperceptible ($\ell_\infty\geq40/255$).

\begin{table}[H]
\centering
\resizebox{\columnwidth}{!}{%
\begin{tabular}{@{}lcccccc@{}}
\toprule
\textbf{Dataset} & \textbf{\makecell{Seen \\ ACC}} & \textbf{\makecell{Seen \\ ASR}} & \textbf{\makecell{Unseen \\ ACC}} & \textbf{\makecell{Unseen \\ ASR}} & \textbf{\makecell{ACC \\ Diff.}} & \textbf{\makecell{ASR \\ Diff.}} \\ \midrule
\multicolumn{1}{l|}{\textbf{Caltech101}} & 98.19\% & \multicolumn{1}{c|}{99.94\%} & 91.48\% & \multicolumn{1}{c|}{\textbf{94.00\%}} & 6.71\% & 5.94\% \\
\multicolumn{1}{l|}{\textbf{DTD}} & 73.43\% & \multicolumn{1}{c|}{98.91\%} & 45.37\% & \multicolumn{1}{c|}{\textbf{95.72\%}} & 28.06\% & 3.19\% \\
\multicolumn{1}{l|}{\textbf{EuroSAT}} & 92.31\% & \multicolumn{1}{c|}{100.00\%} & 49.33\% & \multicolumn{1}{c|}{\textbf{100.00\%}} & 42.98\% & 0.00\% \\
\multicolumn{1}{l|}{\textbf{FGVC\_Aircraft}} & 38.96\% & \multicolumn{1}{c|}{99.58\%} & 28.13\% & \multicolumn{1}{c|}{\textbf{98.44\%}} & 10.83\% & 1.14\% \\
\multicolumn{1}{l|}{\textbf{Food101}} & 88.93\% & \multicolumn{1}{c|}{99.93\%} & 86.48\% & \multicolumn{1}{c|}{\textbf{99.82\%}} & 2.45\% & 0.11\% \\
\multicolumn{1}{l|}{\textbf{ImageNet}} & 74.97\% & \multicolumn{1}{c|}{99.54\%} & 65.90\% & \multicolumn{1}{c|}{98.25\%} & 9.07\% & 1.29\% \\
\multicolumn{1}{l|}{\textbf{Flowers102}} & 96.11\% & \multicolumn{1}{c|}{100.00\%} & 71.13\% & \multicolumn{1}{c|}{\textbf{100.00\%}} & 24.98\% & 0.00\% \\
\multicolumn{1}{l|}{\textbf{OxfordPets}} & 93.66\% & \multicolumn{1}{c|}{99.83\%} & 96.18\% & \multicolumn{1}{c|}{\textbf{92.47\%}} & -2.52\% & 7.36\% \\
\multicolumn{1}{l|}{\textbf{SUN397}} & 77.89\% & \multicolumn{1}{c|}{99.60\%} & 71.21\% & \multicolumn{1}{c|}{\textbf{98.55\%}} & 6.68\% & 1.05\% \\
\multicolumn{1}{l|}{\textbf{UCF101}} & 82.37\% & \multicolumn{1}{c|}{99.95\%} & 65.57\% & \multicolumn{1}{c|}{\textbf{99.26\%}} & 16.80\% & 0.69\% \\ \midrule
\textbf{Average} & \textbf{81.68\%} & \textbf{99.73\%} & \textbf{67.08\%} & \textbf{97.65\%} & \textbf{14.60\%} & \textbf{2.08\%} \\ \bottomrule
\end{tabular}%
}
\caption{ACC and ASR of Blended attack models on Seen and Unseen class subsets.}
\label{tab:acc_asr_stats_blended}
\end{table}

\subsection{SIBA attack performance}
SIBA stands for \textbf{S}parse and \textbf{I}nvisible \textbf{A}ttack. Their aim is to create a trigger that is simultaneously sparse ($\ell_0=1600$) and imperceptible ($\ell_\infty=8/255$). The attack achieves a low ASR due to the perturbation being restricted to only 3\% of total pixels, due to which the meta-net is unable to receive a strong backdoor signal. It's worth noting that increasing the $\ell_0$ bound beyond 1600 would increase ASR but violate the sparsity constraint.

\begin{table}[H]
\centering
\resizebox{\columnwidth}{!}{%
\begin{tabular}{@{}lcccccc@{}}
\toprule
\textbf{Dataset} & \textbf{\makecell{Seen \\ ACC}} & \textbf{\makecell{Seen \\ ASR}} & \textbf{\makecell{Unseen \\ ACC}} & \textbf{\makecell{Unseen \\ ASR}} & \textbf{\makecell{ACC \\ Diff.}} & \textbf{\makecell{ASR \\ Diff.}} \\ \midrule
\multicolumn{1}{l|}{\textbf{Caltech101}} & 95.03\% & \multicolumn{1}{c|}{63.52\%} & 92.79\% & \multicolumn{1}{c|}{\textbf{38.32\%}} & 2.24\% & 25.20\% \\
\multicolumn{1}{l|}{\textbf{DTD}} & 61.47\% & \multicolumn{1}{c|}{82.49\%} & 48.26\% & \multicolumn{1}{c|}{\textbf{71.99\%}} & 13.21\% & 10.50\% \\
\multicolumn{1}{l|}{\textbf{EuroSAT}} & 81.40\% & \multicolumn{1}{c|}{94.67\%} & 44.31\% & \multicolumn{1}{c|}{\textbf{63.90\%}} & 37.09\% & 30.77\% \\
\multicolumn{1}{l|}{\textbf{FGVC\_Aircraft}} & 30.85\% & \multicolumn{1}{c|}{57.86\%} & 32.99\% & \multicolumn{1}{c|}{\textbf{13.38\%}} & -2.14\% & 44.48\% \\
\multicolumn{1}{l|}{\textbf{Food101}} & 88.75\% & \multicolumn{1}{c|}{78.09\%} & 89.90\% & \multicolumn{1}{c|}{\textbf{73.80\%}} & -1.15\% & 4.29\% \\
\multicolumn{1}{l|}{\textbf{ImageNet}} & 71.32\% & \multicolumn{1}{c|}{19.90\%} & 66.51\% & \multicolumn{1}{c|}{13.97\%} & 4.81\% & 5.93\% \\
\multicolumn{1}{l|}{\textbf{Flowers102}} & 89.08\% & \multicolumn{1}{c|}{93.83\%} & 73.05\% & \multicolumn{1}{c|}{\textbf{92.13\%}} & 16.03\% & 1.70\% \\
\multicolumn{1}{l|}{\textbf{OxfordPets}} & 86.98\% & \multicolumn{1}{c|}{48.04\%} & 85.00\% & \multicolumn{1}{c|}{\textbf{42.55\%}} & 1.98\% & 5.49\% \\
\multicolumn{1}{l|}{\textbf{SUN397}} & 72.09\% & \multicolumn{1}{c|}{30.48\%} & 70.26\% & \multicolumn{1}{c|}{\textbf{29.48\%}} & 1.83\% & 1.00\% \\
\multicolumn{1}{l|}{\textbf{UCF101}} & 82.11\% & \multicolumn{1}{c|}{94.14\%} & 72.72\% & \multicolumn{1}{c|}{\textbf{82.31\%}} & 9.39\% & 11.83\% \\ \midrule
\textbf{Average} & \textbf{75.91\%} & \textbf{66.30\%} & \textbf{67.58\%} & \textbf{52.18\%} & \textbf{8.33\%} & \textbf{14.12\%} \\ \bottomrule
\end{tabular}%
}
\caption{SIBA attack ACC and ASR values on seen and unseen subsets for each dataset.}
\label{tab:acc_asr_stats_siba}
\end{table}

\subsection{WaNet attack performance}
WaNet or warping-based backdoor distorts the entire image using a geometric warp grid. The warp shifts pixel positions rather than adding noise, making the change imperceptible. The training occurs in three different modes: (i) clean — clean image with original label, (ii) attack — image distorted via backdoor warping paired with backdoor label, and (iii) noise — image distorted via randomly perturbed backdoor warping paired with original label. For each training image, the mode is selected with probabilities $p_{normal}=0.7, p_{attack}=0.1$, and $p_{noise}=0.2$. 

\begin{figure*}
    \centering
    \includegraphics[width=1\textwidth]{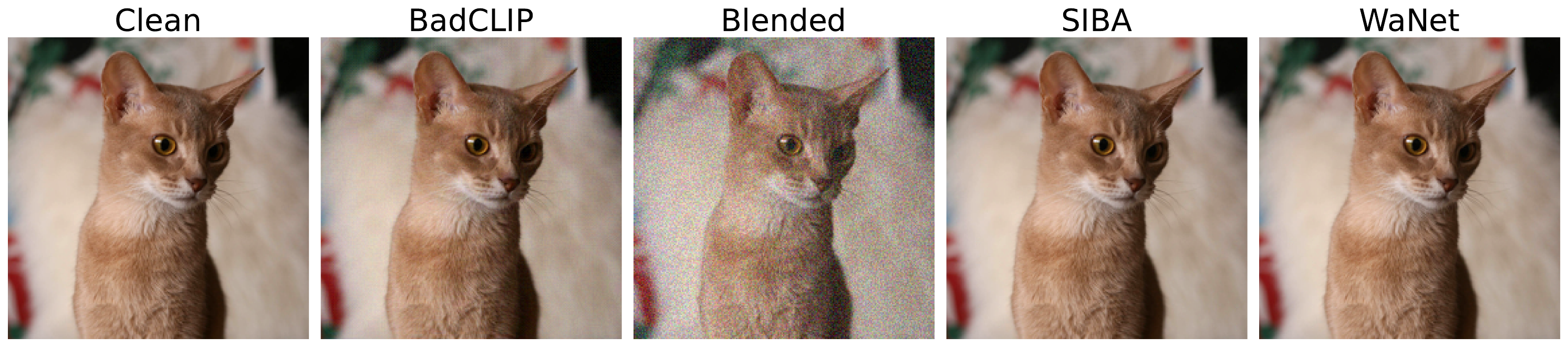}
    \caption{Clean and Triggered image pairs for each attack type. All triggers are visually imperceptible except Blended.}
    \label{fig:trigger_vis_oxpets}
\end{figure*}

\begin{table}[H]
\centering
\resizebox{\columnwidth}{!}{%
\begin{tabular}{@{}lcccccc@{}}
\toprule
\textbf{Dataset} & \textbf{\makecell{Seen \\ ACC}} & \textbf{\makecell{Seen \\ ASR}} & \textbf{\makecell{Unseen \\ ACC}} & \textbf{\makecell{Unseen \\ ASR}} & \textbf{\makecell{ACC \\ Diff.}} & \textbf{\makecell{ASR \\ Diff.}} \\ \midrule
\multicolumn{1}{l|}{\textbf{Caltech101}} & 97.87\% & \multicolumn{1}{c|}{95.67\%} & 90.94\% & \multicolumn{1}{c|}{\textbf{87.66\%}} & 6.93\% & 8.01\% \\
\multicolumn{1}{l|}{\textbf{DTD}} & 75.48\% & \multicolumn{1}{c|}{80.07\%} & 39.47\% & \multicolumn{1}{c|}{\textbf{75.12\%}} & 36.01\% & 4.95\% \\
\multicolumn{1}{l|}{\textbf{EuroSAT}} & 92.71\% & \multicolumn{1}{c|}{86.79\%} & 43.49\% & \multicolumn{1}{c|}{\textbf{44.36\%}} & 49.22\% & 42.43\% \\
\multicolumn{1}{l|}{\textbf{FGVC\_Aircraft}} & 38.24\% & \multicolumn{1}{c|}{98.68\%} & 31.07\% & \multicolumn{1}{c|}{\textbf{75.52\%}} & 7.17\% & 23.16\% \\
\multicolumn{1}{l|}{\textbf{Food101}} & 89.17\% & \multicolumn{1}{c|}{96.83\%} & 88.91\% & \multicolumn{1}{c|}{\textbf{95.20\%}} & 0.26\% & 1.63\% \\
\multicolumn{1}{l|}{\textbf{ImageNet}} & 74.76\% & \multicolumn{1}{c|}{96.32\%} & 66.38\% & \multicolumn{1}{c|}{\textbf{90.02\%}} & 8.38\% & 6.30\% \\
\multicolumn{1}{l|}{\textbf{Flowers102}} & 95.16\% & \multicolumn{1}{c|}{97.53\%} & 70.21\% & \multicolumn{1}{c|}{\textbf{98.44\%}} & 24.95\% & -0.91\% \\
\multicolumn{1}{l|}{\textbf{OxfordPets}} & 93.49\% & \multicolumn{1}{c|}{81.99\%} & 94.81\% & \multicolumn{1}{c|}{\textbf{64.39\%}} & -1.32\% & 17.60\% \\
\multicolumn{1}{l|}{\textbf{SUN397}} & 78.76\% & \multicolumn{1}{c|}{97.26\%} & 71.62\% & \multicolumn{1}{c|}{\textbf{96.39\%}} & 7.14\% & 0.87\% \\
\multicolumn{1}{l|}{\textbf{UCF101}} & 85.17\% & \multicolumn{1}{c|}{94.04\%} & 68.91\% & \multicolumn{1}{c|}{\textbf{89.30\%}} & 16.26\% & 4.74\% \\ \midrule
\textbf{Average} & \textbf{82.08\%} & \textbf{92.52\%} & \textbf{66.58\%} & \textbf{81.64\%} & \textbf{15.50\%} & \textbf{10.88\%} \\ \bottomrule
\end{tabular}%
}
\caption{WaNet attack ACC and ASR values on seen and unseen subsets for each dataset.}
\label{tab:acc_asr_stats_wanet}
\end{table}
The triggered images for each attack type are visualized in Figure \ref{fig:trigger_vis_oxpets}.



\section{Adaptive Attack Against CLIP-Inspector: BadCLIP Adaptive}
\label{sec:adaptive_badclip}

In the main paper, we introduce \textbf{BadCLIP Adaptive}, a two-phase variant of BadCLIP designed to make the backdoor highly specific, such that only a single, exact trigger pattern should activate the target class, whereas small perturbations around this trigger should revert to the clean label. Here, we outline the training procedure.

\subsection{Phase 1: Standard BadCLIP}

Phase~1 mirrors the original BadCLIP's trigger-aware prompt learning method.
Let $\tilde{p}(y=i \mid x)$ denote the prompt-tuned classifier's
posterior and $t$ the attacker's target class. BadCLIP optimizes a
backdoor (trigger) loss
\begin{equation}
\mathcal{L}_{\text{tri}}(\theta, \delta)
= \mathbb{E}_{x_i}
\big[
  - \log \tilde{p}(y = t \mid x_i + \delta)
\big],
\end{equation}
together with a clean classification loss
\begin{equation}
\mathcal{L}_{\text{cle}}(\theta)
= \mathbb{E}_{(x_i,y_i)}
\big[
  - \log \tilde{p}(y = y_i \mid x_i)
\big],
\end{equation}
subject to an $\ell_\infty$ budget
$\|\delta\|_\infty \le \epsilon$ in the normalized image space.
The total Phase~1 loss is
\begin{equation}
\mathcal{L}_{\text{total}}^{(1)}(\theta, \delta)
= \mathcal{L}_{\text{tri}}(\theta, \delta)
+ \mathcal{L}_{\text{cle}}(\theta),
\end{equation}
and we jointly update the model parameters $\theta$ and the trigger
$\delta$ using SGD. After convergence we obtain an optimized trigger
$\delta^\star$ and freeze it as
\begin{equation}
\delta_{\text{fixed}} = \delta^\star.
\end{equation}

\subsection{Phase 2: Specificity Fine-Tuning}

Phase~2 keeps $\delta_{\text{fixed}}$ frozen and further trains the prompt-tuned model to \emph{reject} perturbed versions of the trigger. Let $\epsilon$ denote the $\ell_\infty$ budget and $\sigma$ the per-channel normalization scale (from CLIP preprocessing). We generate a perturbed trigger by adding Gaussian noise and clipping back to the admissible
range:
\begin{equation}
\begin{aligned}
\eta &\sim
\mathcal{N}\big(0, (\alpha \epsilon / \sigma)^2 I\big), \\
\delta_{\text{perturbed}}
&= \operatorname{clip}\big(
    \delta_{\text{fixed}} + \eta,\,
    -\epsilon/\sigma,\,
    \epsilon/\sigma
  \big),
\end{aligned}
\end{equation}
where $\alpha$ is the perturbation strength and the clipping is applied elementwise in the normalized space.

For each training image $x_i$ with label $y_i$, Phase 2 uses three types of
inputs:
\begin{itemize}
  \item Clean images $x_i$ with label $y_i$;
  \item Exact-trigger images $x_i + \delta_{\text{fixed}}$ with target
        label $t$;
  \item Perturbed-trigger images $x_i + \delta_{\text{perturbed}}$ with
        the \emph{clean} label $y_i$.
\end{itemize}

We can equivalently write the Phase~2 objective as a sum of three losses:
\begin{equation}
\begin{aligned}
\mathcal{L}_{\text{cle}}(\theta)
&= \mathbb{E}_{(x_i, y_i)}
   \big[
     - \log \tilde{p}(y_i \mid x_i)
   \big], \\
\mathcal{L}_{\text{tri}}(\theta)
&= \mathbb{E}_{x_i}
   \big[
     - \log \tilde{p}(t \mid x_i + \delta_{\text{fixed}})
   \big], \\
\mathcal{L}_{\text{spec}}(\theta)
&= \mathbb{E}_{(x_i, y_i)}
   \big[
     - \log \tilde{p}(y_i \mid x_i + \delta_{\text{perturbed}})
   \big].
\end{aligned}
\end{equation}
The Phase~2 loss is then
\begin{equation}
\mathcal{L}_{\text{total}}^{(2)}(\theta)
= \mathcal{L}_{\text{cle}}(\theta)
+ \mathcal{L}_{\text{tri}}(\theta)
+ \lambda_{\text{spec}}\, \mathcal{L}_{\text{spec}}(\theta),
\end{equation}
In other words, Phase~2 encourages
\begin{itemize}
  \item high confidence on the target class $t$ for images stamped with
        the \emph{exact} trigger $\delta_{\text{fixed}}$, and
  \item clean predictions for images stamped with \emph{perturbed}
        triggers $\delta_{\text{perturbed}}$.
\end{itemize}
This forces the model to associate only the exact trigger pattern with
the target label, while nearby perturbations are pushed back to the
original class. As a result, the ``basin of attraction'' around
$\delta_{\text{fixed}}$ becomes narrower in trigger space: random
$\ell_\infty$-bounded perturbations that approximate the trigger are
much less likely to activate the backdoor, which reduces the ASR of
approximate triggers such as those reconstructed by CLIP-Inspector.
In our implementation we set $\alpha=0.5$ and $\lambda_{\text{spec}} = 1$.

Empirically, we observe that enforcing high specificity inevitably lowers the ASR of the BadCLIP optimised trigger. Small deviations introduced by clean pixels no longer steer images reliably toward the target class. Despite this reduced ASR, our method still finds the narrow shortcut in 8 of 10 backdoored models. (On UCF101, the reconstructed trigger achieves \(<50\%\) ASR, so its anomaly score is considered invalid.) Table \ref{tab:adaptive_attack} shows the seen vs unseen class metrics for each dataset, along with the reconstructed trigger ASR and overall anomaly score metrics for our method. These results show that any adaptation to reduce detection success would inevitably harm the attacker's objective, proving the effectiveness of our method against adaptive attackers.

\begin{table}[H]
\centering
\resizebox{\columnwidth}{!}{%
\begin{tabular}{@{}lcccccc@{}}
\toprule
\textbf{Dataset} &
\textbf{\begin{tabular}[c]{@{}c@{}}Seen\\ACC\end{tabular}} &
\textbf{\begin{tabular}[c]{@{}c@{}}Seen\\ASR\end{tabular}} &
\textbf{\begin{tabular}[c]{@{}c@{}}Unseen\\ACC\end{tabular}} &
\textbf{\begin{tabular}[c]{@{}c@{}}Unseen\\ASR\end{tabular}} &
\textbf{\begin{tabular}[c]{@{}c@{}}CI\\ASR\end{tabular}} &
\textbf{\begin{tabular}[c]{@{}c@{}}Anomaly\\Score\end{tabular}} \\
\midrule
\multicolumn{1}{l|}{Caltech101}  & 97.29\% & \multicolumn{1}{c|}{64.11\%} & 93.45\% & \multicolumn{1}{c|}{58.95\%} & 67.29\% & 11.20 \\
\multicolumn{1}{l|}{DTD}         & 76.93\% & \multicolumn{1}{c|}{63.04\%} & 47.34\% & \multicolumn{1}{c|}{56.25\%} & 31.54\% &  6.86 \\
\multicolumn{1}{l|}{EuroSAT}     & 90.50\% & \multicolumn{1}{c|}{52.38\%} & 46.56\% & \multicolumn{1}{c|}{20.18\%} & 67.77\% &  1.66 \\
\multicolumn{1}{l|}{FGVC\_Aircraft} & 34.75\% & \multicolumn{1}{c|}{93.04\%} & 34.07\% & \multicolumn{1}{c|}{35.03\%} & 58.50\% &  6.32 \\
\multicolumn{1}{l|}{Food101}     & 90.22\% & \multicolumn{1}{c|}{56.35\%} & 90.84\% & \multicolumn{1}{c|}{48.38\%} & 95.02\% & 13.32 \\
\multicolumn{1}{l|}{ImageNet}    &  75.94\% & \multicolumn{1}{c|}{ 71.68\%} &  69.68\% & \multicolumn{1}{c|}{ 52.93\%} &  98.83\% &  11.67 \\
\multicolumn{1}{l|}{Flowers102}  & 93.83\% & \multicolumn{1}{c|}{69.33\%} & 71.77\% & \multicolumn{1}{c|}{65.39\%} & 94.04\% & 11.91 \\
\multicolumn{1}{l|}{OxfordPets}  & 93.15\% & \multicolumn{1}{c|}{37.26\%} & 93.75\% & \multicolumn{1}{c|}{21.57\%} & 56.54\% & 10.62 \\
\multicolumn{1}{l|}{SUN397}      & 79.01\% & \multicolumn{1}{c|}{67.73\%} & 75.89\% & \multicolumn{1}{c|}{60.32\%} & 97.17\% & 12.04 \\
\multicolumn{1}{l|}{UCF101}      & 84.54\% & \multicolumn{1}{c|}{65.80\%} & 69.60\% & \multicolumn{1}{c|}{61.18\%} & 30.96\% &  8.97 \\
\midrule
\textbf{Average} & \textbf{81.62\%} & \textbf{64.07\%} & \textbf{69.30\%} & \textbf{48.02\%} & \textbf{69.77\%} & \textbf{9.45} \\
\bottomrule
\end{tabular}%
}
\caption{BadCLIP-adaptive results. Enforcing trigger specificity sharply reduces ASR, yet the behavioural anomaly metric of CLIP-Inspector still flags most backdoored models.}
\label{tab:adaptive_attack}
\end{table}

\section{Anomaly score discrimination metrics}
In this section, we highlight the metrics used by each detection method to mark anomalous classes. 

\subsection{CI}
Our approach flags a class as anomalous when, during a single-epoch optimisation, it exhibits both (i) an unusually low average reconstruction loss and (ii) a high attack-success rate (ASR) for the recovered trigger. A sharp drop in loss within one epoch indicates a “shortcut” in the loss landscape leading directly to the target class. This shortcut exists only for the backdoor target class and not for any of the other classes.  
Table \ref{tab:ci_disc_metrics} shows these metrics for clean and backdoored models for the Caltech101 dataset. The difference columns show their deviation from the maximum value amongst non-target classes. For clean models, the ASR value is lower than other classes, while the loss average is high. The opposite is true for backdoored models, confirming the presence of backdoors.

\begin{table}[H]
\centering
\resizebox{\columnwidth}{!}{%
\begin{tabular}{@{}lcccc@{}}
\toprule
 & \multicolumn{4}{c}{\textbf{CLIP-Inspector}} \\ \cmidrule(l){2-5} 
\textbf{Attack} & \textbf{\begin{tabular}[c]{@{}c@{}}Backdoor \\ ASR\end{tabular}} & \textbf{\begin{tabular}[c]{@{}c@{}}ASR\\ Difference\end{tabular}} & \textbf{\begin{tabular}[c]{@{}c@{}}Backdoor\\ Loss Average\end{tabular}} & \textbf{\begin{tabular}[c]{@{}c@{}}Loss Average\\ Difference\end{tabular}} \\ \midrule
\multicolumn{1}{l|}{\textbf{Clean}} & 7.51 & \multicolumn{1}{c|}{-26.39} & 5.2218 & -2.0343 \\
\multicolumn{1}{l|}{\textbf{BadCLIP}} & 94.74 & \multicolumn{1}{c|}{66.27} & -8.2344 & -15.7907 \\
\multicolumn{1}{l|}{\textbf{Blended}} & 99.02 & \multicolumn{1}{c|}{72.68} & -11.2263 & -20.8507 \\
\multicolumn{1}{l|}{\textbf{SIBA}} & 92.69 & \multicolumn{1}{c|}{65.75} & -6.2807 & -13.1315 \\
\multicolumn{1}{l|}{\textbf{WaNet}} & 73.33 & \multicolumn{1}{c|}{50.61} & -4.5774 & -12.1891 \\ \bottomrule
\end{tabular}%
}
\caption{ASR and Average Optimization loss values for triggers reconstructed via CLIP-Inspector for Caltech101 dataset.}
\label{tab:ci_disc_metrics}
\end{table}

\subsection{NC}
Neural Cleanse marks anomalous classes based on the reconstructed trigger's mask size or $\ell_1$ norm. Their method is directly correlated to their sparsity assumption, as a sparse trigger would have a low $\ell_1$ norm. However, NC is not able to differentiate between clean and backdoored models based on the `shortcut' behaviour trend we discussed earlier. NC behaves similarly for both clean and backdoored classes, creating triggers with high ASR for every class due to their dynamic regularization scheme. 

\begin{table}[H]
\centering
\resizebox{\columnwidth}{!}{%
\begin{tabular}{@{}lcccc@{}}
\toprule
 & \multicolumn{4}{c}{\textbf{Neural Cleanse}} \\ \cmidrule(l){2-5} 
\textbf{Attack} & \textbf{\begin{tabular}[c]{@{}c@{}}Backdoor \\ ASR\end{tabular}} & \textbf{\begin{tabular}[c]{@{}c@{}}ASR\\ Difference\end{tabular}} & \textbf{\begin{tabular}[c]{@{}c@{}}Backdoor\\ Mask Size\end{tabular}} & \textbf{\begin{tabular}[c]{@{}c@{}}Mask Size\\ Difference\end{tabular}} \\ \midrule
\multicolumn{1}{l|}{\textbf{Clean}} & 99.64 & \multicolumn{1}{c|}{-0.29} & 9874.8175 & -7479.3773 \\
\multicolumn{1}{l|}{\textbf{BadCLIP}} & 99.49 & \multicolumn{1}{c|}{-0.5} & 3836.3224 & -13693.3142 \\
\multicolumn{1}{l|}{\textbf{Blended}} & 99.34 & \multicolumn{1}{c|}{-0.61} & 574.5364 & -19629.362 \\
\multicolumn{1}{l|}{\textbf{SIBA}} & 99.52 & \multicolumn{1}{c|}{-0.47} & 2861.0645 & -14295.5531 \\
\multicolumn{1}{l|}{\textbf{WaNet}} & 99.48 & \multicolumn{1}{c|}{-0.5} & 4895.8796 & -13152.8507 \\ \bottomrule
\end{tabular}%
}
\caption{ASR and Average Optimization loss values for triggers reconstructed via Neural Cleanse.}
\label{tab:nc_disc_metrics}
\end{table}

\subsection{PixB}
Pixel Backdoor uses perturbed pixel counts instead of mask sizes to mark anomalous classes. Its behaviour is similar to Neural Cleanse as it makes similar sparsity assumptions and is unable to differentiate between clean and backdoor models in a clear manner.

\begin{table}[H]
\centering
\resizebox{\columnwidth}{!}{%
\begin{tabular}{@{}lcccc@{}}
\toprule
 & \multicolumn{4}{c}{\textbf{Pixel Backdoor}} \\ \cmidrule(l){2-5} 
\textbf{Attack} & \textbf{\begin{tabular}[c]{@{}c@{}}Backdoor \\ ASR\end{tabular}} & \textbf{\begin{tabular}[c]{@{}c@{}}ASR\\ Difference\end{tabular}} & \textbf{\begin{tabular}[c]{@{}c@{}}Backdoor\\ Loss Average\end{tabular}} & \textbf{\begin{tabular}[c]{@{}c@{}}Loss Average\\ Difference\end{tabular}} \\ \midrule
\multicolumn{1}{l|}{\textbf{Clean}} & 85.77 & \multicolumn{1}{c|}{-5.8} & 64122.1 & -36842.2 \\
\multicolumn{1}{l|}{\textbf{BadCLIP}} & 88.64 & \multicolumn{1}{c|}{-3.77} & 35026 & -61933.6 \\
\multicolumn{1}{l|}{\textbf{Blended}} & 97.08 & \multicolumn{1}{c|}{7.4} & 5455.2 & -104308.3 \\
\multicolumn{1}{l|}{\textbf{SIBA}} & 91.18 & \multicolumn{1}{c|}{-1.74} & 25955.1 & -65539 \\
\multicolumn{1}{l|}{\textbf{WaNet}} & 87.08 & \multicolumn{1}{c|}{-5.08} & 57801.5 & -44050.9 \\ \bottomrule
\end{tabular}%
}
\caption{ASR and Average Optimization loss values for triggers reconstructed via Pixel Backdoor.}
\label{tab:pixb_disc_metrics}
\end{table}

\begin{figure}
    \centering
    \includegraphics[width=\linewidth]{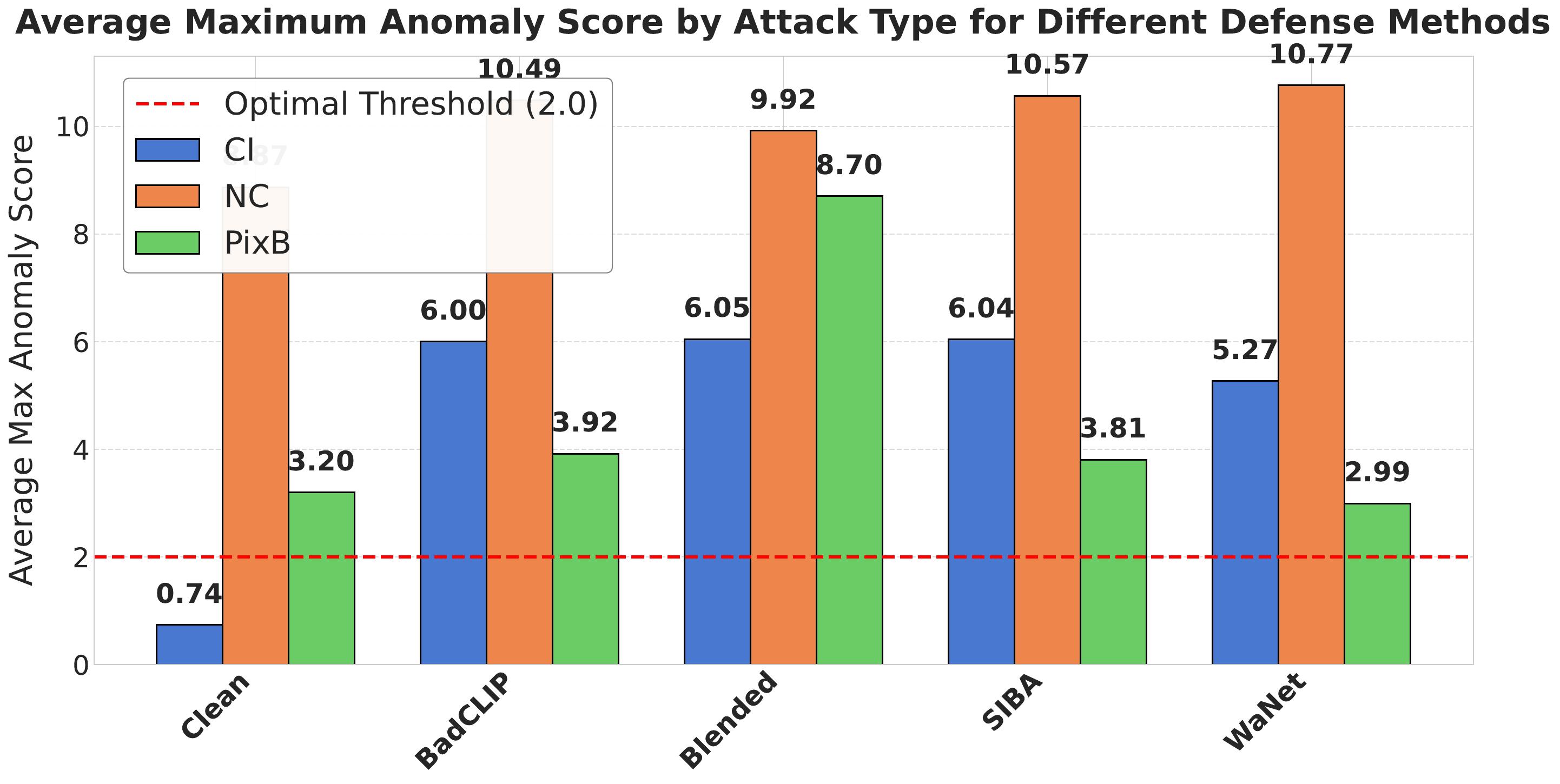}
    \caption{Average anomaly scores for each method averaged across 10 datasets for each attack type. CI shows clear distinction between clean and backdoored models while other methods fail to do so.}
    \label{fig:anomaly_score_avg}
\end{figure}

Overall anomaly scores averaged across 10 datasets for each attack type and clean models are given in Figure \ref{fig:anomaly_score_avg}. Only CLIP-Inspector showcases a low anomaly score for clean models and a high anomaly score for backdoored models. The scores for other methods are not discriminatory at all, resulting in a high number of false positives and false negatives.

\section{Ablation study}
\subsection{Varying number of samples in OOD dataset}
We vary the number of samples used for backdoor detection in BadCLIP models from 100 to 500 and 1000 samples. The ASR of the reconstructed trigger for the backdoor class is shown in Table \ref{tab:sample_vary}. ASR drops sharply when sample count falls from 500 to 100. However, the drop observed when reducing samples from 1000 to 500 is not significant for most datasets except the UCF101 dataset, where the ASR drops by 30\%.

\begin{table}[btp]
\centering
\small
\resizebox{0.8\columnwidth}{!}{%
\begin{tabular}{@{}lccc@{}}
\toprule
 & \multicolumn{3}{c}{\textbf{Num OOD samples}} \\ \cmidrule(l){2-4} 
\textbf{Dataset} & \textbf{100} & \textbf{500} & \textbf{1000} \\ \midrule
\multicolumn{1}{l|}{\textbf{Caltech101}} & 58\% & 86.40\% & 98.20\% \\
\multicolumn{1}{l|}{\textbf{DTD}} & 11\% & 73.60\% & 80.40\% \\
\multicolumn{1}{l|}{\textbf{EuroSAT}} & 92\% & 97.60\% & 99.60\% \\
\multicolumn{1}{l|}{\textbf{FGVC\_Aircraft}} & 58\% & 83.40\% & 94.30\% \\
\multicolumn{1}{l|}{\textbf{Food101}} & 73\% & 100\% & 99.80\% \\
\multicolumn{1}{l|}{\textbf{ImageNet}} & 100\% & 100\% & 99.70\% \\
\multicolumn{1}{l|}{\textbf{Flowers102}} & 15\% & 100\% & 100\% \\
\multicolumn{1}{l|}{\textbf{OxfordPets}} & 96\% & 98.60\% & 97.20\% \\
\multicolumn{1}{l|}{\textbf{SUN397}} & 92\% & 99\% & 99.90\% \\
\multicolumn{1}{l|}{\textbf{UCF101}} & 1\% & 53.20\% & 83.70\% \\ \midrule
\multicolumn{1}{l|}{\textbf{Average}} & 60\% & 89.18\% & 95.28\% \\ \bottomrule
\end{tabular}%
}
\caption{ASR of CI’s reconstructed trigger on the backdoor class when varying the number of samples used for backdoor detection in BadCLIP models. }
\label{tab:sample_vary}
\end{table}

\subsection{Varying batch size $|B|$}
As we run CLIP-Inspector for only one epoch, varying the batch size plays an important role as it directly correlates with the number of times the trigger is optimized by the Adam optimizer. We vary the batch size and show the ASR of the reconstructed trigger for the backdoor target class in Table \ref{tab:batch_size_vary} when using 1000 samples for trigger inversion. A batch size of 1 yields noisy gradient updates, whereas a batch size of 64 results in only 16 optimization steps, which are too few for convergence. We therefore adopt a batch size of 32 (32 optimization steps) for all our experiments. 

\begin{table}[btp]
\centering
\small
\resizebox{0.8\columnwidth}{!}{%
\begin{tabular}{@{}lccc@{}}
\toprule
 & \multicolumn{3}{c}{\textbf{Batch Size}} \\ \cmidrule(l){2-4} 
\textbf{Dataset} & \textbf{1} & \textbf{32} & \textbf{64} \\ \midrule
\multicolumn{1}{l|}{\textbf{Caltech101}} & 69.9\% & 98.20\% & 91.3\% \\
\multicolumn{1}{l|}{\textbf{DTD}} & 32.2\% & 80.40\% & 87.4\% \\
\multicolumn{1}{l|}{\textbf{EuroSAT}} & 98.1\% & 99.60\% & 99.8\% \\
\multicolumn{1}{l|}{\textbf{FGVC\_Aircraft}} & 52.6\% & 94.30\% & 89.2\% \\
\multicolumn{1}{l|}{\textbf{Food101}} & 96.5\% & 99.80\% & 100.0\% \\
\multicolumn{1}{l|}{\textbf{ImageNet}} & 99.7\% & 99.70\% & 100.0\% \\
\multicolumn{1}{l|}{\textbf{Flowers102}} & 84.5\% & 100\% & 100.0\% \\
\multicolumn{1}{l|}{\textbf{OxfordPets}} & 90.2\% & 97.20\% & 98.9\% \\
\multicolumn{1}{l|}{\textbf{SUN397}} & 91.1\% & 99.90\% & 99.2\% \\
\multicolumn{1}{l|}{\textbf{UCF101}} & 44.0\% & 83.70\% & 18.7\% \\ \midrule
\multicolumn{1}{l|}{\textbf{Average}} & 75.9\% & 95.3\% & 88.5\% \\ \bottomrule
\end{tabular}%
}
\caption{ASR values for the backdoor target class when varying the batch size used for trigger inversion (1000 OOD samples, 1 epoch).}
\label{tab:batch_size_vary}
\end{table}

\subsection{ID vs OOD sample selection}
Switching to In-distribution (ID) samples from Out-of-Distribution (OOD) samples for trigger inversion has little effect on reconstructed trigger ASR values. This shows that ID samples are not necessary to create an effective trigger, owing to the strong cross-domain generalization capability of the BadCLIP trigger. Results are in Table \ref{tab:id_vs_ood}.

\begin{table}[btp]
\centering
\resizebox{0.7\columnwidth}{!}{%
\begin{tabular}{@{}lcc@{}}
\toprule
 & \multicolumn{2}{c}{\textbf{ID vs OOD samples}} \\ \cmidrule(l){2-3} 
\textbf{Dataset} & \textbf{ID} & \textbf{OOD} \\ \midrule
\multicolumn{1}{l|}{\textbf{Caltech101}} & 99.7\% & 98.20\% \\
\multicolumn{1}{l|}{\textbf{DTD}} & 89.6\% & 80.40\% \\
\multicolumn{1}{l|}{\textbf{EuroSAT}} & 98.8\% & 99.60\% \\
\multicolumn{1}{l|}{\textbf{FGVC\_Aircraft}} & 99.2\% & 94.30\% \\
\multicolumn{1}{l|}{\textbf{Food101}} & 100.0\% & 99.80\% \\
\multicolumn{1}{l|}{\textbf{ImageNet}} & 100.0\% & 99.70\% \\
\multicolumn{1}{l|}{\textbf{Flowers102}} & 99.7\% & 100\% \\
\multicolumn{1}{l|}{\textbf{OxfordPets}} & 99.5\% & 97.20\% \\
\multicolumn{1}{l|}{\textbf{SUN397}} & 100.0\% & 99.90\% \\
\multicolumn{1}{l|}{\textbf{UCF101}} & 99.0\% & 83.70\% \\ \midrule
\multicolumn{1}{l|}{\textbf{Average}} & 98.6\% & 95.3\% \\ \bottomrule
\end{tabular}%
}
\caption{ASR values for the backdoor target class when using ID vs OOD data for trigger inversion. ASR is reported for the backdoor target class for the BadCLIP attack models.}
\label{tab:id_vs_ood}
\end{table}

\section{Generalization to Encoder-Level Backdoors (No Meta-Net)}
We use the Blended poisoning method to poison the image encoder with three patterns (Gaussian noise, triangle pattern, written text). CI is able to separate clean from poisoned models and identify the target class without altering the inversion process or hyperparameters. Results are given in Table \ref{tab:encoder-level}.
\begin{table}[H]
\centering\small
\begin{tabular}{lccc}
\toprule
Pattern & ASR & CI-ASR & \makecell{Anomaly \\ Score} \\
\midrule
Gaussian & 99.7\% & 98.75\% & 6.15\\
Triangle & 86.1\% & 96.88\% & 5.93\\
Text & 94.7\% & 62.4\% & 4.38\\
\bottomrule
\end{tabular}
\caption{Generality beyond prompt-tuning.}
\label{tab:encoder-level}
\end{table}

\section{Repair Study: Controls and Hyperparameters}
We compare CI-trigger repair against controls: Clean-only FT, Random-$\delta$, Wrong-class $\delta$; and report ACC/ASR values averaged over all datasets for each attack type. Clean models are not considered in this ablation; therefore, the average values may differ from those presented in the main paper. Per-attack metrics for each ablation are reported in Tables \ref{tab:cleanonly_removal}-\ref{tab:nontarget_removal} with overall averages summarized in Table \ref{tab:removal_effectiveness}.
\begin{table}[h]
\centering
\small
\resizebox{\columnwidth}{!}{%
\begin{tabular}{lcccc}
\toprule
\textbf{Condition} & \textbf{ACC before} & \textbf{ACC after} & \textbf{ASR before} & \textbf{ASR after} \\
\midrule
\multicolumn{1}{l|}{Clean-only} & 80.6\% & \multicolumn{1}{l|}{81.4\%} & 84.4\% & 64.6\%\\
\multicolumn{1}{l|}{Random-$\delta$} & 80.6\% & \multicolumn{1}{l|}{81.0\%} & 84.4\% & 52.7\%\\
\multicolumn{1}{l|}{Wrong-class-$\delta$} & 80.6\% & \multicolumn{1}{l|}{80.3\%} & 84.4\% & 47.0\%\\
\multicolumn{1}{l|}{CI-trigger(Ours)} & 80.6\% & \multicolumn{1}{l|}{80.8\%} & 84.4\% & \textbf{6.1\%} \\
\bottomrule
\end{tabular}%
}
\caption{Measuring backdoor removal effectiveness by comparing against different trigger initializations.}
\label{tab:removal_effectiveness}
\end{table}

\begin{table}[t]
\centering
\resizebox{\columnwidth}{!}{%
\begin{tabular}{@{}lcccc@{}}
\toprule
\textbf{Attack}                       & \textbf{ACC before} & \textbf{ACC after}                    & \textbf{ASR before} & \textbf{ASR after} \\ \midrule
\multicolumn{1}{l|}{\textbf{BadCLIP}}           & 81.8\%    & \multicolumn{1}{c|}{81.3\%} & 99.0\%    & 85.2\%    \\
\multicolumn{1}{l|}{\textbf{BadCLIP\_adaptive}} & 81.6\%    & \multicolumn{1}{c|}{81.5\%} & 64.1\%    & 59.4\%    \\
\multicolumn{1}{l|}{\textbf{Blended}}           & 81.7\%    & \multicolumn{1}{c|}{81.8\%} & 99.7\%    & 87.3\%    \\
\multicolumn{1}{l|}{\textbf{SIBA}}              & 75.9\%    & \multicolumn{1}{c|}{79.1\%} & 66.3\%    & 8.3\%    \\
\multicolumn{1}{l|}{\textbf{WaNet}}             & 82.1\%    & \multicolumn{1}{c|}{83.2\%} & 92.7\%    & 82.6\%    \\ \midrule
\multicolumn{1}{l|}{\textbf{Average}}           & 80.6\%    & 81.4\%                      & 84.4\%    & 64.6\%    \\ \bottomrule
\end{tabular}%
}
\caption{Using only a clean dataset for fine tuning does not remove the backdoor.}
\label{tab:cleanonly_removal}
\end{table}

\begin{table}[t]
\centering
\resizebox{\columnwidth}{!}{%
\begin{tabular}{@{}lcccc@{}}
\toprule
\textbf{Attack}                       & \textbf{ACC before} & \textbf{ACC after}                    & \textbf{ASR before} & \textbf{ASR after} \\ \midrule
\multicolumn{1}{l|}{\textbf{BadCLIP}}             & 81.8\%    & \multicolumn{1}{c|}{81.4\%} & 99.0\%    & 7.6\%    \\
\multicolumn{1}{l|}{\textbf{BadCLIP\_adaptive}}   & 81.6\%    & \multicolumn{1}{c|}{81.0\%} & 64.1\%    & 4.2\%    \\
\multicolumn{1}{l|}{\textbf{Blended}}             & 81.7\%    & \multicolumn{1}{c|}{81.7\%} & 99.7\%    & 7.6\%    \\
\multicolumn{1}{l|}{\textbf{SIBA}}                & 75.9\%    & \multicolumn{1}{c|}{77.9\%} & 66.3\%    & 5.8\%    \\
\multicolumn{1}{l|}{\textbf{WaNet}}               & 82.1\%    & \multicolumn{1}{c|}{82.1\%} & 92.7\%    & 5.3\%    \\ \midrule
\multicolumn{1}{l|}{\textbf{Average}}             & 80.6\%    & 80.8\%                      & 84.4\%    & 6.1\%    \\ \bottomrule
\end{tabular}%
}
\caption{Removal Using CI-trigger.}
\label{tab:ci_removal}
\end{table}

\begin{table}[t]
\centering
\resizebox{\columnwidth}{!}{%
\begin{tabular}{@{}lcccc@{}}
\toprule
\textbf{Attack}                       & \textbf{ACC before} & \textbf{ACC after}                    & \textbf{ASR before} & \textbf{ASR after} \\ \midrule
\multicolumn{1}{l|}{\textbf{BadCLIP}}           & 81.8\%    & \multicolumn{1}{c|}{81.1\%} & 99.0\%    & 73.3\%    \\
\multicolumn{1}{l|}{\textbf{BadCLIP\_adaptive}} & 81.6\%    & \multicolumn{1}{c|}{82.0\%} & 64.1\%    & 55.1\% \\
\multicolumn{1}{l|}{\textbf{Blended}}           & 81.7\%    & \multicolumn{1}{c|}{81.5\%} & 99.7\%    & 40.0\%    \\
\multicolumn{1}{l|}{\textbf{SIBA}}              & 75.9\%    & \multicolumn{1}{c|}{78.8\%} & 66.3\%    & 8.6\%    \\
\multicolumn{1}{l|}{\textbf{WaNet}}             & 82.1\%    & \multicolumn{1}{c|}{81.7\%} & 92.7\%    & 86.5\%    \\ \midrule
\multicolumn{1}{l|}{\textbf{Average}}           & 80.6\%    & 81.0\%                      & 84.4\%    & 52.7\%    \\ \bottomrule
\end{tabular}%
}
\caption{Removal using random noise is highly ineffective for all attack types. }
\label{tab:random_removal}
\end{table}

\begin{table}[t]
\centering
\resizebox{\columnwidth}{!}{%
\begin{tabular}{@{}lcccc@{}}
\toprule
\textbf{Attack}                       & \textbf{ACC before} & \textbf{ACC after}                    & \textbf{ASR before} & \textbf{ASR after} \\ \midrule
\multicolumn{1}{l|}{\textbf{BadCLIP}}           & 81.8\%    & \multicolumn{1}{c|}{80.5\%} & 99.0\%    & 60.3\%    \\
\multicolumn{1}{l|}{\textbf{BadCLIP\_adaptive}} & 81.6\%    & \multicolumn{1}{c|}{80.2\%} & 64.1\%    & 43.5\% \\
\multicolumn{1}{l|}{\textbf{Blended}}           & 81.7\%    & \multicolumn{1}{c|}{80.9\%} & 99.7\%    & 56.0\%    \\
\multicolumn{1}{l|}{\textbf{SIBA}}              & 75.9\%    & \multicolumn{1}{c|}{77.8\%} & 66.3\%    & 5.4\%    \\
\multicolumn{1}{l|}{\textbf{WaNet}}             & 82.1\%    & \multicolumn{1}{c|}{82.0\%} & 92.7\%    & 70.0\%    \\ \midrule
\multicolumn{1}{l|}{\textbf{Average}}           & 80.6\%    & 80.3\%                      & 84.4\%    & 47.0\%    \\ \bottomrule
\end{tabular}%
}
\caption{Removal Using CI-inverted trigger from a non-target class.}
\label{tab:nontarget_removal}
\end{table}

\begin{table*}
\centering
\resizebox{\textwidth}{!}{%
\begin{tabular}{@{}lcccccccc@{}}
\toprule
\textbf{Dataset} & \multicolumn{1}{l}{\textbf{\begin{tabular}[c]{@{}l@{}}BadCLIP\\ Trigger \\ SSIM (mean)\end{tabular}}} & \multicolumn{1}{l}{\textbf{\begin{tabular}[c]{@{}l@{}}BadCLIP\\ Trigger \\ SSIM (std)\end{tabular}}} & \multicolumn{1}{l}{\textbf{\begin{tabular}[c]{@{}l@{}}CI Recon.\\ Trigger \\ SSIM (mean)\end{tabular}}} & \multicolumn{1}{l}{\textbf{\begin{tabular}[c]{@{}l@{}}CI Recon. \\ Trigger \\ SSIM (std)\end{tabular}}} & \multicolumn{1}{l}{\textbf{\begin{tabular}[c]{@{}l@{}}NC Recon. \\ Trigger\\ SSIM (mean)\end{tabular}}} & \multicolumn{1}{l}{\textbf{\begin{tabular}[c]{@{}l@{}}NC Recon. \\ Trigger \\ SSIM (std)\end{tabular}}} & \multicolumn{1}{l}{\textbf{\begin{tabular}[c]{@{}l@{}}PixB Recon.\\ Trigger \\ SSIM (mean)\end{tabular}}} & \multicolumn{1}{l}{\textbf{\begin{tabular}[c]{@{}l@{}}PixB Recon. \\ Trigger \\ SSIM (std)\end{tabular}}} \\ \midrule
\multicolumn{1}{l|}{\textbf{Caltech101}} & 0.9641 & \multicolumn{1}{c|}{0.0202} & 0.9373 & \multicolumn{1}{c|}{0.0346} & 0.4414 & \multicolumn{1}{c|}{0.0999} & 0.4355 & 0.1261 \\
\multicolumn{1}{l|}{\textbf{DTD}} & 0.9681 & \multicolumn{1}{c|}{0.0178} & 0.9344 & \multicolumn{1}{c|}{0.0360} & 0.9196 & \multicolumn{1}{c|}{0.0375} & 0.6102 & 0.1106 \\
\multicolumn{1}{l|}{\textbf{EuroSAT}} & 0.9703 & \multicolumn{1}{c|}{0.0165} & 0.9367 & \multicolumn{1}{c|}{0.0350} & 0.9999 & \multicolumn{1}{c|}{0.0001} & 0.6389 & 0.1067 \\
\multicolumn{1}{l|}{\textbf{FGVC\_Aircraft}} & 0.9632 & \multicolumn{1}{c|}{0.0204} & 0.9362 & \multicolumn{1}{c|}{0.0352} & 0.7920 & \multicolumn{1}{c|}{0.0657} & 0.7617 & 0.0813 \\
\multicolumn{1}{l|}{\textbf{Food101}} & 0.9617 & \multicolumn{1}{c|}{0.0215} & 0.9356 & \multicolumn{1}{c|}{0.0354} & 0.5132 & \multicolumn{1}{c|}{0.0894} & 0.7840 & 0.0760 \\
\multicolumn{1}{l|}{\textbf{ImageNet}} & 0.9549 & \multicolumn{1}{c|}{0.0255} & 0.9395 & \multicolumn{1}{c|}{0.0337} & 0.8603 & \multicolumn{1}{c|}{0.0509} & 0.8176 & 0.0669 \\
\multicolumn{1}{l|}{\textbf{Flowers102}} & 0.9608 & \multicolumn{1}{c|}{0.0215} & 0.9348 & \multicolumn{1}{c|}{0.0357} & 0.9655 & \multicolumn{1}{c|}{0.0198} & 0.6289 & 0.1112 \\
\multicolumn{1}{l|}{\textbf{OxfordPets}} & 0.9655 & \multicolumn{1}{c|}{0.0192} & 0.9368 & \multicolumn{1}{c|}{0.0350} & 0.4264 & \multicolumn{1}{c|}{0.1106} & 0.4533 & 0.1251 \\
\multicolumn{1}{l|}{\textbf{SUN397}} & 0.9585 & \multicolumn{1}{c|}{0.0235} & 0.9368 & \multicolumn{1}{c|}{0.0352} & 0.7855 & \multicolumn{1}{c|}{0.0873} & 0.7981 & 0.0739 \\
\multicolumn{1}{l|}{\textbf{UCF101}} & 0.9638 & \multicolumn{1}{c|}{0.0200} & 0.9382 & \multicolumn{1}{c|}{0.0341} & 0.9517 & \multicolumn{1}{c|}{0.0293} & 0.3724 & 0.1214 \\ \midrule
\textbf{Average} & \textbf{0.96309} & 0.02061 & \textbf{0.93663} & 0.03499 & \textbf{0.76556} & 0.05905 & \textbf{0.63006} & 0.09992 \\ \bottomrule
\end{tabular}%
}
\caption{SSIM values for the original BadCLIP trigger and triggers reconstructed using our method and baselines. Our method achieves an SSIM value of 0.93, close to the original trigger's value of 0.96.}
\label{tab:SSIM_badclip}
\end{table*}

\section{Structural Similarity (SSIM) scores for original and reconstructed triggers}
We show the Structural Similarity or SSIM values for the original and reconstructed triggers (for the backdoor target class) in this section. Recall that the BadCLIP trigger is imperceptible and pervasive, the Blended trigger is pervasive but not imperceptible, SIBA is sparse and imperceptible, and the warping distortion of WaNet can also be termed imperceptible and pervasive. Thus, BadCLIP, SIBA, and WaNet have high SSIM scores ($>0.9$) as compared to the Blended attack ($\approx0.5$). Our detection method CI uses an imperceptibility threshold or $\mathcal{L}_\infty$ of 4/255, leading to an average SSIM score of 0.93. On the other hand, NC and PixB have no such imperceptibility constraint. SSIM is computed between clean and triggered images over the OOD pool.  

\subsection{BadCLIP}
The BadCLIP trigger is highly imperceptible with $\mathcal{L}_\infty = 4/255$, leading to a high average SSIM score of 0.96 across all datasets. The SSIM values for reconstructed triggers, inverted from BadCLIP models, are given in Table \ref{tab:SSIM_badclip}.

\subsection{Blended}
The Blended trigger consists of uniform random noise with normalized pixel values in the range $[0,1]$. The trigger is thus highly visible and spread across every pixel, which is what leads to a low SSIM score of 0.505. Kindly refer Table \ref{tab:SSIM_blended} for SSIM values for reconstructed triggers of each defense method. 
\begin{table*}[]
\centering
\resizebox{\textwidth}{!}{%
\begin{tabular}{@{}lcccccccc@{}}
\toprule
\textbf{Dataset} & \multicolumn{1}{l}{\textbf{\begin{tabular}[c]{@{}l@{}}Blended\\ Trigger \\ SSIM (mean)\end{tabular}}} & \multicolumn{1}{l}{\textbf{\begin{tabular}[c]{@{}l@{}}Blended\\ Trigger \\ SSIM (std)\end{tabular}}} & \multicolumn{1}{l}{\textbf{\begin{tabular}[c]{@{}l@{}}CI Recon.\\ Trigger \\ SSIM (mean)\end{tabular}}} & \multicolumn{1}{l}{\textbf{\begin{tabular}[c]{@{}l@{}}CI Recon. \\ Trigger \\ SSIM (std)\end{tabular}}} & \multicolumn{1}{l}{\textbf{\begin{tabular}[c]{@{}l@{}}NC Recon. \\ Trigger\\ SSIM (mean)\end{tabular}}} & \multicolumn{1}{l}{\textbf{\begin{tabular}[c]{@{}l@{}}NC Recon. \\ Trigger \\ SSIM (std)\end{tabular}}} & \multicolumn{1}{l}{\textbf{\begin{tabular}[c]{@{}l@{}}PixB Recon.\\ Trigger \\ SSIM (mean)\end{tabular}}} & \multicolumn{1}{l}{\textbf{\begin{tabular}[c]{@{}l@{}}PixB Recon. \\ Trigger \\ SSIM (std)\end{tabular}}} \\ \midrule
\multicolumn{1}{l|}{\textbf{Caltech101}} & 0.5742 & \multicolumn{1}{c|}{0.146} & 0.9344 & \multicolumn{1}{c|}{0.0361} & 0.9414 & \multicolumn{1}{c|}{0.0263} & 0.9927 & 0.0078 \\
\multicolumn{1}{l|}{\textbf{DTD}} & 0.5475 & \multicolumn{1}{c|}{0.152} & 0.9346 & \multicolumn{1}{c|}{0.0357} & 0.7717 & \multicolumn{1}{c|}{0.0372} & 0.841 & 0.0593 \\
\multicolumn{1}{l|}{\textbf{EuroSAT}} & 0.5347 & \multicolumn{1}{c|}{0.152} & 0.9332 & \multicolumn{1}{c|}{0.0365} & 0.9859 & \multicolumn{1}{c|}{0.0131} & 0.7639 & 0.0815 \\
\multicolumn{1}{l|}{\textbf{FGVC\_Aircraft}} & 0.4375 & \multicolumn{1}{c|}{0.145} & 0.9345 & \multicolumn{1}{c|}{0.0359} & 0.9405 & \multicolumn{1}{c|}{0.0287} & 0.7141 & 0.0936 \\
\multicolumn{1}{l|}{\textbf{Food101}} & 0.523 & \multicolumn{1}{c|}{0.144} & 0.9352 & \multicolumn{1}{c|}{0.0356} & 0.9999 & \multicolumn{1}{c|}{0.0001} & 0.9294 & 0.0307 \\
\multicolumn{1}{l|}{\textbf{ImageNet}} & 0.5446 & \multicolumn{1}{c|}{0.143} & 0.9346 & \multicolumn{1}{c|}{0.0357} & 0.9948 & \multicolumn{1}{c|}{0.006} & 0.8365 & 0.0639 \\
\multicolumn{1}{l|}{\textbf{Flowers102}} & 0.463 & \multicolumn{1}{c|}{0.145} & 0.9349 & \multicolumn{1}{c|}{0.0358} & 0.9837 & \multicolumn{1}{c|}{0.0133} & 0.8772 & 0.0495 \\
\multicolumn{1}{l|}{\textbf{OxfordPets}} & 0.4857 & \multicolumn{1}{c|}{0.146} & 0.9349 & \multicolumn{1}{c|}{0.0361} & 0.9141 & \multicolumn{1}{c|}{0.0546} & 0.9661 & 0.0178 \\
\multicolumn{1}{l|}{\textbf{SUN397}} & 0.5296 & \multicolumn{1}{c|}{0.147} & 0.9345 & \multicolumn{1}{c|}{0.0358} & 0.9237 & \multicolumn{1}{c|}{0.0182} & 0.885 & 0.0472 \\
\multicolumn{1}{l|}{\textbf{UCF101}} & 0.4347 & \multicolumn{1}{c|}{0.150} & 0.9339 & \multicolumn{1}{c|}{0.036} & 0.9877 & \multicolumn{1}{c|}{0.0092} & 0.9741 & 0.0146 \\ \midrule
\textbf{Average} & \textbf{0.5057} & 0.147 & \textbf{0.93447} & 0.03592 & \textbf{0.94434} & 0.02067 & \textbf{0.878} & 0.04659 \\ \bottomrule
\end{tabular}%
}
\caption{SSIM values for the original Blended trigger and triggers reconstructed using our method and baselines.}
\label{tab:SSIM_blended}
\end{table*}

\subsection{SIBA}
Because SIBA is sparse and imperceptible, per-pixel perturbation is extremely low, leading to a high SSIM score of 0.99. Refer Table \ref{tab:SSIM_siba} for the SSIM scores of reconstructed triggers. 

\begin{table*}
\centering
\resizebox{\textwidth}{!}{%
\begin{tabular}{@{}lcccccccc@{}}
\toprule
\textbf{Dataset} & \multicolumn{1}{l}{\textbf{\begin{tabular}[c]{@{}l@{}}SIBA\\ Trigger \\ SSIM (mean)\end{tabular}}} & \multicolumn{1}{l}{\textbf{\begin{tabular}[c]{@{}l@{}}SIBA\\ Trigger \\ SSIM (std)\end{tabular}}} & \multicolumn{1}{l}{\textbf{\begin{tabular}[c]{@{}l@{}}CI Recon.\\ Trigger \\ SSIM (mean)\end{tabular}}} & \multicolumn{1}{l}{\textbf{\begin{tabular}[c]{@{}l@{}}CI Recon. \\ Trigger \\ SSIM (std)\end{tabular}}} & \multicolumn{1}{l}{\textbf{\begin{tabular}[c]{@{}l@{}}NC Recon. \\ Trigger\\ SSIM (mean)\end{tabular}}} & \multicolumn{1}{l}{\textbf{\begin{tabular}[c]{@{}l@{}}NC Recon. \\ Trigger \\ SSIM (std)\end{tabular}}} & \multicolumn{1}{l}{\textbf{\begin{tabular}[c]{@{}l@{}}PixB Recon.\\ Trigger \\ SSIM (mean)\end{tabular}}} & \multicolumn{1}{l}{\textbf{\begin{tabular}[c]{@{}l@{}}PixB Recon. \\ Trigger \\ SSIM (std)\end{tabular}}} \\ \midrule
\multicolumn{1}{l|}{\textbf{Caltech101}} & 0.9995 & \multicolumn{1}{c|}{0.0003} & 0.9367 & \multicolumn{1}{c|}{0.0348} & 0.6516 & \multicolumn{1}{c|}{0.1067} & 0.7586 & 0.0855 \\
\multicolumn{1}{l|}{\textbf{DTD}} & 0.9995 & \multicolumn{1}{c|}{0.0002} & 0.9375 & \multicolumn{1}{c|}{0.0345} & 0.864 & \multicolumn{1}{c|}{0.0391} & 0.6125 & 0.1116 \\
\multicolumn{1}{l|}{\textbf{EuroSAT}} & 0.9995 & \multicolumn{1}{c|}{0.0002} & 0.9374 & \multicolumn{1}{c|}{0.0348} & 0.9781 & \multicolumn{1}{c|}{0.0192} & 0.5037 & 0.1208 \\
\multicolumn{1}{l|}{\textbf{FGVC\_Aircraft}} & 0.9994 & \multicolumn{1}{c|}{0.0003} & 0.9378 & \multicolumn{1}{c|}{0.0341} & 0.5587 & \multicolumn{1}{c|}{0.0896} & 0.5822 & 0.1156 \\
\multicolumn{1}{l|}{\textbf{Food101}} & 0.9996 & \multicolumn{1}{c|}{0.0002} & 0.9364 & \multicolumn{1}{c|}{0.0351} & 0.8724 & \multicolumn{1}{c|}{0.0665} & 0.499 & 0.1242 \\
\multicolumn{1}{l|}{\textbf{ImageNet}} & 0.9995 & \multicolumn{1}{c|}{0.0003} & 0.9372 & \multicolumn{1}{c|}{0.0344} & 0.9631 & \multicolumn{1}{c|}{0.0251} & 0.7942 & 0.0726 \\
\multicolumn{1}{l|}{\textbf{Flowers102}} & 0.9995 & \multicolumn{1}{c|}{0.0003} & 0.9379 & \multicolumn{1}{c|}{0.0341} & 0.9245 & \multicolumn{1}{c|}{0.0321} & 0.5887 & 0.1164 \\
\multicolumn{1}{l|}{\textbf{OxfordPets}} & 0.9996 & \multicolumn{1}{c|}{0.0002} & 0.9383 & \multicolumn{1}{c|}{0.034} & 0.9859 & \multicolumn{1}{c|}{0.0105} & 0.6971 & 0.0958 \\
\multicolumn{1}{l|}{\textbf{SUN397}} & 0.9996 & \multicolumn{1}{c|}{0.0002} & 0.9352 & \multicolumn{1}{c|}{0.0354} & 0.8904 & \multicolumn{1}{c|}{0.0432} & 0.6498 & 0.1064 \\
\multicolumn{1}{l|}{\textbf{UCF101}} & 0.9995 & \multicolumn{1}{c|}{0.0003} & 0.9395 & \multicolumn{1}{c|}{0.0339} & 0.3683 & \multicolumn{1}{c|}{0.0894} & 0.613 & 0.1133 \\ \midrule
\textbf{Average} & \textbf{0.99952} & 0.00025 & \textbf{0.93739} & 0.03451 & \textbf{0.8057} & 0.05214 & \textbf{0.62988} & 0.10622 \\ \bottomrule
\end{tabular}%
}
\caption{SSIM values for the original SIBA trigger and triggers reconstructed using our method and baselines. SIBA trigger is highly sparse and imperceptible, evident by the 0.99 SSIM score.}
\label{tab:SSIM_siba}
\end{table*}

\subsection{WaNet}
The WaNet trigger is a smooth geometric warp derived from a $k\times k$ control grid and scaled by a strength factor $s$, subtly shifting every pixel’s position. We choose $k=4$ and $s=0.1$ to generate a highly imperceptible trigger with an SSIM value of $\approx$0.94 across datasets. Please refer to Table \ref{tab:SSIM_wanet} for SSIM scores of reconstructed triggers. 

\begin{table*}
\centering
\resizebox{\textwidth}{!}{%
\begin{tabular}{@{}lcccccccc@{}}
\toprule
\textbf{Dataset} & \multicolumn{1}{l}{\textbf{\begin{tabular}[c]{@{}l@{}}WaNet\\ Trigger \\ SSIM (mean)\end{tabular}}} & \multicolumn{1}{l}{\textbf{\begin{tabular}[c]{@{}l@{}}WaNet\\ Trigger \\ SSIM (std)\end{tabular}}} & \multicolumn{1}{l}{\textbf{\begin{tabular}[c]{@{}l@{}}CI Recon.\\ Trigger \\ SSIM (mean)\end{tabular}}} & \multicolumn{1}{l}{\textbf{\begin{tabular}[c]{@{}l@{}}CI Recon. \\ Trigger \\ SSIM (std)\end{tabular}}} & \multicolumn{1}{l}{\textbf{\begin{tabular}[c]{@{}l@{}}NC Recon. \\ Trigger\\ SSIM (mean)\end{tabular}}} & \multicolumn{1}{l}{\textbf{\begin{tabular}[c]{@{}l@{}}NC Recon. \\ Trigger \\ SSIM (std)\end{tabular}}} & \multicolumn{1}{l}{\textbf{\begin{tabular}[c]{@{}l@{}}PixB Recon.\\ Trigger \\ SSIM (mean)\end{tabular}}} & \multicolumn{1}{l}{\textbf{\begin{tabular}[c]{@{}l@{}}PixB Recon. \\ Trigger \\ SSIM (std)\end{tabular}}} \\ \midrule
\multicolumn{1}{l|}{\textbf{Caltech101}} & 0.9312 & \multicolumn{1}{c|}{0.0186} & 0.9358 & \multicolumn{1}{c|}{0.0354} & 0.4515 & \multicolumn{1}{c|}{0.1005} & 0.3998 & 0.1233 \\
\multicolumn{1}{l|}{\textbf{DTD}} & 0.9342 & \multicolumn{1}{c|}{0.0186} & 0.9372 & \multicolumn{1}{c|}{0.0347} & 0.7839 & \multicolumn{1}{c|}{0.0608} & 0.4233 & 0.1229 \\
\multicolumn{1}{l|}{\textbf{EuroSAT}} & 0.9309 & \multicolumn{1}{c|}{0.0186} & 0.9374 & \multicolumn{1}{c|}{0.0344} & 0.9588 & \multicolumn{1}{c|}{0.0207} & 0.3643 & 0.119 \\
\multicolumn{1}{l|}{\textbf{FGVC\_Aircraft}} & 0.9233 & \multicolumn{1}{c|}{0.0184} & 0.9373 & \multicolumn{1}{c|}{0.0345} & 0.6507 & \multicolumn{1}{c|}{0.1058} & 0.3748 & 0.1213 \\
\multicolumn{1}{l|}{\textbf{Food101}} & 0.9473 & \multicolumn{1}{c|}{0.0189} & 0.9376 & \multicolumn{1}{c|}{0.0346} & 0.8571 & \multicolumn{1}{c|}{0.0648} & 0.6744 & 0.1013 \\
\multicolumn{1}{l|}{\textbf{ImageNet}} & 0.9512 & \multicolumn{1}{c|}{0.0190} & 0.9356 & \multicolumn{1}{c|}{0.0354} & 0.784 & \multicolumn{1}{c|}{0.0798} & 0.3707 & 0.1211 \\
\multicolumn{1}{l|}{\textbf{Flowers102}} & 0.9277 & \multicolumn{1}{c|}{0.0185} & 0.9355 & \multicolumn{1}{c|}{0.0356} & 0.7949 & \multicolumn{1}{c|}{0.0586} & 0.399 & 0.1235 \\
\multicolumn{1}{l|}{\textbf{OxfordPets}} & 0.9592 & \multicolumn{1}{c|}{0.0191} & 0.936 & \multicolumn{1}{c|}{0.0352} & 0.7422 & \multicolumn{1}{c|}{0.09} & 0.7226 & 0.0902 \\
\multicolumn{1}{l|}{\textbf{SUN397}} & 0.9342 & \multicolumn{1}{c|}{0.0186} & 0.9356 & \multicolumn{1}{c|}{0.0356} & 0.8134 & \multicolumn{1}{c|}{0.0749} & 0.5623 & 0.1176 \\
\multicolumn{1}{l|}{\textbf{UCF101}} & 0.9421 & \multicolumn{1}{c|}{0.0188} & 0.9375 & \multicolumn{1}{c|}{0.0346} & 0.5323 & \multicolumn{1}{c|}{0.1079} & 0.4257 & 0.1231 \\ \midrule
\textbf{Average} & \textbf{0.9381} & 0.0187 & \textbf{0.93655} & 0.035 & \textbf{0.73688} & 0.07638 & \textbf{0.47169} & 0.11633 \\ \bottomrule
\end{tabular}%
}
\caption{SSIM values for the original WaNet trigger and triggers reconstructed using our method and baselines. We apply a $k=4$ control-grid distortion with noise strength $s=0.1$.
}
\label{tab:SSIM_wanet}
\end{table*}
\end{document}